\documentclass[aps,prd,showpacs,twocolumn,superscriptaddress,nofootinbib,preprintnumbers]{revtex4-1} 

\makeatletter
\def\p@subsection{}
\makeatother

\usepackage{color,graphicx,amsmath,amssymb,lipsum}
\usepackage[colorlinks=true,citecolor=blue,linkcolor=blue,urlcolor=blue, backref=false,pdfborder={0 0 0}]{hyperref}

\usepackage[normalem]{ulem}
\usepackage[utf8]{inputenc} 
\usepackage{mathtools}

\usepackage{float}

\usepackage{multirow}
\usepackage{ulem}
\usepackage{diagbox}
\usepackage{tabularx}

\usepackage[dvipsnames]{xcolor}
\usepackage{mathrsfs}

\newcommand{\be}{\begin{equation}}
\newcommand{\ee}{\end{equation}}
\newcommand{\beqa}{\begin{eqnarray}}
\newcommand{\eeqa}{\end{eqnarray}}

\renewcommand\L{\Lambda}

\newcommand{\bseq}{\begin{subequations}}
\newcommand{\eseq}{\end{subequations}}

\newcommand{\sh}{{\rm shot}}

\newcommand{\kmax}{k_{\rm max}}

\newcommand{\hMpc}{\,h \text{Mpc}^{-1}}
\newcommand{\Mpc}{\text{Mpc}}

\newcommand{\Gpch}{h^{-1}\text{Gpc}}
\newcommand{\Mpch}{\text{Mpc}/h}

\newcommand{\N}{\mathcal N}
\newcommand{\eff}{{\rm eff}}
\newcommand{\shot}{{\rm shot}}

\definecolor{MyDarkOrange}{rgb}{0.8, 0.4, 0.0}

\def\gsim{\raise0.3ex\hbox{$\;>$\kern-0.75em\raise-1.1ex\hbox{$\sim\;$}}}
\def\lsim{\raise0.3ex\hbox{$\;<$\kern-0.75em\raise-1.1ex\hbox{$\sim\;$}}}

\def\beqn#1{\begin{equation}\label{#1}}
\def\eeqn{\end{equation}}

\def\beqa#1{\begin{eqnarray}\label{#1}}
\def\eeqa{\end{eqnarray}}

\def\Z2{$\mathcal{Z_2}$}

\newcommand {\ignore}[1]{}

\begin{document}

\preprint{MIT-CTP/5797}
% % \preprint{TTK-20-11}

\title{
On priors and scale cuts in EFT-based full-shape 
analyses
}

\author{Anton Chudaykin}
\email{anton.chudaykin@unige.ch}
\affiliation{D\'epartement de Physique Th\'eorique and Center for Astroparticle Physics,\\
Universit\'e de Gen\`eve, 24 quai Ernest  Ansermet, 1211 Gen\`eve 4, Switzerland}
\author{Mikhail M. Ivanov}
\email{ivanov99@mit.edu}
\affiliation{Center for Theoretical Physics -- a Leinweber Institute, Massachusetts Institute of Technology, 
Cambridge, MA 02139, USA}
\author{Takahiro Nishimichi}
\email{takahiro.nishimichi@yukawa.kyoto-u.ac.jp}
\affiliation{Department of Astrophysics and Atmospheric Sciences, Faculty of Science, Kyoto Sangyo University, Motoyama, Kamigamo, Kita-ku, Kyoto 603-8555, Japan}
\affiliation{Center for Gravitational Physics and Quantum Information, Yukawa Institute for Theoretical Physics, Kyoto University, Kyoto 606-8502, Japan}
\affiliation{Kavli Institute for the Physics and Mathematics of the Universe (WPI), The University of Tokyo Institutes for Advanced Study (UTIAS), The University of Tokyo, Kashiwa, Chiba 277-8583, Japan}

\begin{abstract} 
Parameter estimation from galaxy survey data from the full-shape method depends on scale cuts and priors on EFT parameters. 
The effects of priors, including the so-called 
``prior volume'' phenomenon
have been originally 
studied in Ivanov et al. (2019) 
and subsequent works.
In this note, 
we repeat 
and extend 
these tests and also apply them to other 
priors used 
in the literature.
We point out that in addition to the 
``prior volume'' effect there is a more dangerous
effect that is largely overlooked: a systematic bias on 
cosmological parameters due to overoptimistic 
scale cuts. Unlike the ``prior volume'' effect, 
this is a genuine systematic bias 
due to two-loop corrections that 
does not vanish 
with better priors or 
with larger data volumes. 
Our study is based on the high fidelity 
BOSS-like PT Challenge simulation data which offer many advantages over analyses
based on 
synthetic data generated with fitting pipelines.
We show that some analysis choices
associated with the PyBird code,
especially the
scale cuts, significantly 
bias parameter recovery, overestimating $\sigma_8$ by over $5\%$ (equivalent to $1\sigma$). 
The bias on measured EFT parameters is even more significant. 
In contrast, 
the analysis choices associated with the
CLASS-PT code lead to much smaller ($\lesssim 1\%$) shifts in cosmological parameters
based on their best-fit values. 
\end{abstract}

\maketitle

\section{Introduction}

Effective-field theory (EFT) based full shape
(FS)
analysis is a new tool to 
extract fundamental cosmological parameters from the shape 
of the galaxy power spectrum~\cite{Ivanov:2019pdj,DAmico:2019fhj,Philcox:2020vvt,Chen:2021wdi,Philcox:2021kcw,Chudaykin:2022nru}. It relies on 
the EFT of large-scale 
structure~\cite{Baumann:2010tm,Carrasco:2012cv,Ivanov:2022mrd} 
and its variants~\cite{Blas:2015qsi,Blas:2016sfa,Chen:2020fxs,Chen:2020zjt} for the theoretical
modeling of the galaxy clustering data on quasi-linear scales. 
EFT predictions are approximate as the calculations
are carried over only up to a given order of perturbation theory
that matches the required precision. EFT models also involve a number
of free nuisance parameters 
which capture the ignorance about the galaxy formation physics 
on small scales and have to fitted to data alongside cosmological 
parameters. EFT-based full-shape results are thus subject to 
two effects associated with the above limitations. 

Any EFT-based full-shape analysis has to assume 
priors on nuisance parameters in order to ensure that they take physically consistent values. 
Within the Bayesian framework commonly
used in cosmology,
the posterior distribution depends on priors by virtue of Bayes' theorem. 
Even with informative physically motivated priors, there are residual 
degeneracies between the EFT and nuisance parameters.\footnote{This degeneracy between cosmological and bias parameters is frequentest in nature. This degeneracy exists, for instance, in Fisher forecasts like~\cite{Wadekar:2020hax,Cabass:2022epm}, and affect the parameter estimation even in the absence of any prior volume effects. Reducing this 
physical degeneracy is the primary goal of the simulation based prior approach of~\cite{Ivanov:2024hgq,Cabass:2024wob,Ivanov:2024xgb}. } For low precision data, some of these degeneracies may result in non-Gaussian posteriors,
whose 1D marginalized projections are 
shifted w.r.t. the best-fit values. 
This projection effect is often referred to as the ``prior volume'' effect.\footnote{This nomenclature, strictly speaking, is accurate only when the likelihood has flat directions, so the 1D marginalized 
posterior is dominated by volume of the prior rather than the likelihood.}
The distance between the best-fit value and the mean of the 1D marginalized contours
does not have any interpretation in the 
Bayesian framework.
Given this reason, ``prior volume'' effects,
especially at the level of 
$(1-2)\sigma$,
are not often discussed in the literature.\footnote{See e.g.~\cite{OShaughnessy:2013zfw,Biscoveanu:2021eht,Olsen:2021qin} for prior volume
effects in gravitational wave astronomy.}
In EFT-FS analyses, however, 
these effects have recently attracted significant 
attention.

``Prior volume'' effects in the context of the EFT-based 
full-shape analyses were first studied in \cite{Ivanov:2019pdj,Chudaykin:2020ghx,Philcox:2021kcw} for
the analysis based on the \texttt{CLASS-PT} code~\cite{Chudaykin:2020aoj}.~\footnote{The ``prior volume'' effects
in EFT-based full shape analyses
were recently revisited e.g.  in~\cite{Simon:2022lde}.}
In particular, the full-shape analysis of the BOSS data
~\cite{Alam:2016hwk}
in 
\cite{Ivanov:2019pdj} estimated them at the $1\sigma$
level based on a single chunk of the BOSS data.
Refs.~\cite{Ivanov:2019pdj,Chudaykin:2020ghx} pointed out that these effects
strongly depend on the volume of the survey, i.e. the quality of data.
Results of~\cite{Ivanov:2019pdj,Chudaykin:2020ghx}  
suggest that they will not represent a problem for the interpretation of results from future galaxy surveys 
such as Euclid~\cite{Laureijs:2011gra} and DESI~\cite{Aghamousa:2016zmz}
because the statistical error
of these surveys will
be smaller than that of BOSS.
Even for BOSS the
``prior volume'' effects can be mitigated 
in practice in two ways: by including more data (see e.g.~\cite{Ivanov:2023qzb,Chen:2024vuf}), which increases the 
role of the likelihood, or by using better priors informed
from simulations~\cite{Cabass:2024wob,Ivanov:2024hgq,Ivanov:2024xgb,Akitsu:2024lyt}.\footnote{A similar approach was originally put forward in the context of the Halo-Zel'dovich model~\cite{Sullivan:2021sof}.} 

In any realistic analysis ``prior volume'' effects are coupled with 
the bias due to two-loop corrections omitted in commonly used one-loop EFT models. 
Unlike the projection effects, the two-loop corrects represent a genuine theoretical error 
which does not vanish as the quality of data improves. The theoretical error
associated with the two loop corrections has a very steep dependence on the momentum 
scale cut $\kmax$, e.g. for dark matter in real space it scales as
$k^{3.3}$ relative to the linear theory prediction~\cite{Baldauf:2016sjb,Chudaykin:2020hbf}. Thus, a slight increase of $\kmax$  from $0.2~\hMpc$ to $0.25~\hMpc$
increases the error by a factor of 2, and can make it an important source 
of systematic uncertainty. 

A good way to estimate bias due to the theoretical error
on estimated cosmological parameters
is to fit mock data from numerical 
simulations. For accurate estimates
one needs to use simulations whose volume is 
much larger than that of the current surveys. 
An example is given by the PT Challenge Simulation~\cite{Nishimichi:2020tvu}, 
whose cumulative volume of $566~[h^{-1}{\rm Gpc}]^3$
is hundred times larger than the volume of the BOSS survey.
The PT Challenge ``masked'' data were used 
in the past to validate the 
scale cuts in 
EFT-based full-shape analyses.
The outstanding fidelity and 
volume of this simulation
allowed one to detect a two-loop systematic error 
in the BOSS-like data already at $\kmax=0.2~\hMpc$.
The method of community 
challenges that started with the PT Challenge 
has been 
recently extended to other methods 
beyond perturbation theory and beyond
the traditional clustering statistics in 
the Beyond-2pt challenge~\cite{Beyond-2pt:2024mqz}.

The presence of the theory systematic error 
in EFT-FS analyses shows
why parameter projection effects should not be 
considered in isolation.
For instance, one may choose overoptimistic scale cuts 
or priors in their analysis, which would result
in a very small prior volume effect, but both 
1D posteriors and the best-fit values will be 
biased away from the true values. 
As we show here, certain analysis choices associated with the \texttt{PyBird} code~\cite{DAmico:2020kxu} lead to a non-negligible systematic bias in parameter recovery.

Many analyses choices associated with the \texttt{CLASS-PT} code~\cite{Chudaykin:2020aoj}
were validated against the PT Challenge simulation in the past. 
In this note, we repeat and improve these analyses 
by adjusting them to simulate the 
BOSS survey as close as possible.
This includes the first use of the PT Challenge
bispectrum data across the full range of redshifts 
relevant for BOSS. Previous analyses of refs.~\cite{Philcox:2021kcw,Ivanov:2023qzb}
were based on the PT Challenge 
bispectrum at a single redshift $z=0.61$. 
In this work we include in the 
analysis the $z=0.38$
PT Challenge bispectrum as well. 
In addition, we study in detail analyses choices 
associated with the \texttt{PyBird} code, within the same framework.

The use of high-fidelity BOSS-like PT Challenge simulation data
offers many advantages over analyses 
based on synthetic data generated within the fitting pipeline, done e.g. in~\cite{Simon:2022lde}.
First, it enables the estimation of the theory systematic 
bias arising from the imperfect modeling of the target statistics.
Second, it allows one to estimate the quality of the recovery 
of true cosmological parameters.
Third, the large-volume data allow one to easily estimate the true best-fit model
by rescaling the covariance. This method is thus not sensitive 
to noise in the covariance matrices~\cite{Hartlap:2006kj,Sellentin:2015waz,Howlett:2017vwp,Wadekar:2020hax,Philcox:2020zyp} and does not suffer from 
numerical uncertainties associated with the likelihood minimization.

The rest of this note is structured as follows.
Section~\ref{sec:sum}
summarizes our main findings. 
Section~\ref{sec:comp}
compares the analyses choices associated 
with the \texttt{CLASS-PT} and \texttt{PyBird} codes, while 
Section~\ref{sec:ptc}
presents our main analyses of the 
PT Challenge simulation data.

\section{Summary and Main Conclusions}
\label{sec:sum}

We compare analysis choices associated with the \texttt{PyBird} and \texttt{CLASS-PT} EFT-FS likelihoods.   
Following \cite{Nishimichi:2020tvu} we dub those two analysis configurations as “West coast” (WC) and “East coast” (EC) models. We also consider two extensions of these models, EC1/2 designed to highlight
certain analysis aspects.
The EC model that matches the actual BOSS analysis~\cite{Philcox:2021kcw}
is referred to as EC3.
Note that the WC analysis here is the one was used 
in the actual BOSS analysis~\cite{DAmico:2019fhj}.~\footnote{It is different from the WC analysis implemented 
in the PT Challenge~\cite{Nishimichi:2020tvu}, which used a more conservative $k_\mathrm{max}$ value.} 

We analyze mock data simulating the BOSS survey.
For each model (WC/EC), the datavector is kept the same, while 
for the covariance we consider two choices: the BOSS covariance extracted
from the Patchy mocks~\cite{Kitaura:2015uqa}, 
and the Gaussian covariance corresponding to 
a survey with 
100 times the BOSS volume across four BOSS data chunks.
This choice allows us to get the results
free of the parameter projection effects which are equivalent to the best-fit 
values. Our main results are:

\textbf{1. WC model tests.} 
The WC analysis choices result in a $0.8\sigma_{\rm BOSS}$ shift of the posterior mean of $\omega_{cdm}$ and introduce a $\sigma_{\rm BOSS}$ bias in the best-fit value of $\sigma_8$ ($\sigma_{\rm BOSS}$ is the statistical error from the BOSS data).
For $\omega_{cdm}$, this is a coupled
effect of parameter projection and the inconsistency in the 
modeling of the stochastic contributions. 
These two biases can be eliminated by including the 
scale-dependent stochastic contributions in the monopole and choosing a more conservative $k_\mathrm{max}$.

The recovered best-fit value of $\sigma_8$ is shifted up by $\approx 5\%$ as a 
result of the two-loop corrections. 
This bias
remains in the combinations with other datasets like \textit{Planck} CMB.

This WC case shows that the ``prior volume'' metric,
i.e. the distance between 
the best-fit and the marginalized mean value, 
can be misleading
if the scale cuts are not calibrated properly. 
In the WC model the prior volume effect
on $\sigma_8$ is relatively 
small, 
but the best-fit is biased w.r.t. the true value, which impacts parameter recovery.

In addition, the EFT parameter recovery is strongly biased in the WC model.
This raises concerns about the robustness of analyses of the non-local primordial non-Gaussianity (PNG), 
which are sensitive to the EFT bias parameters such as $b_2$ and $b_{\mathcal{G}_2}$~\cite{Cabass:2022wjy,Cabass:2022ymb,DAmico:2022gki,Ivanov:2024hgq}.

\textbf{2. EC model tests.} 
The EC parameter choices
lead to the unbiased
recovery of $h$
and $\omega_{cdm}$
estimated both by best-fits
and the 1D marginalized 
means. 
The best-fit of the mass fluctuation
amplitude $\sigma_8$
is recovered with a $1.5\%$
bias, which is negligible 
given the BOSS errorbars~\cite{Philcox:2021kcw,Chen:2024vuf}. 
The peak of the 1D marginalized
posterior is shifted 
down by about $2\sigma$. 
While this shift somewhat complicates
the interpretation of the 
BOSS-only results, 
this is entirely a prior 
effect, which does not affect
the ability of the  
\texttt{CLASS-PT}-based pipeline to
recover true cosmological
parameters as estimated 
by their best-fit
values.
The combinations of BOSS
FS-EFT with $Planck$
CMB data and other external 
datasets suppress the prior effects~\cite{Ivanov:2019hqk,Chudaykin:2020ghx}. 
Hence, our analysis confirms
the robustness of 
the corresponding results 
obtained with the \texttt{CLASS-PT} based pipeline, e.g.~\cite{Ivanov:2019hqk,Chudaykin:2020ghx,Ivanov:2020ril,Xu:2021rwg,Nunes:2022bhn,Kumar:2022vee,Rubira:2022xhb,Rogers:2023ezo,He:2023dbn,He:2023oke,Camarena:2023cku,McDonough:2023qcu}.
In addition, the prior effects 
can be mitigated by using more 
data as in~\cite{Chen:2024vuf}, 
or simulation-based priors~\cite{Cabass:2024wob,Ivanov:2024hgq,Ivanov:2024xgb,Akitsu:2024lyt}.

\textbf{3. The validity of the single volume approximation and estimates of 
parameter projection effects for DESI/Euclid.}
On the technical side, 
we present a detailed comparison of 
our results obtained
with the full simulation of the 
BOSS data across four data chunks
with the previous tests 
based on a single PT Challenge
redshift with a covariance rescaled to 
match the cumulative volume of BOSS DR12
luminous red galaxy (LRG) sample.
We call the latter approach 
a ``single volume approximation''.

We find that for WC and early 
variants of the EC analyses 
the single volume approximation 
works reasonably well. 
For the most recent \texttt{CLASS-PT} based
analyses using the galaxy bispectrum~\cite{Philcox:2021kcw}, 
however, the single 
volume approximation
underestimates 
the parameter projection
effects. In this case, simulating 
all sub-volumes of the BOSS data 
is necessary to accurately access the 
parameter projection effects. Carrying 
out this analysis consistently 
is one of the 
main results 
of our study.

Finally, our single-volume 
results can be treated as 
an approximation for ongoing stage-IV surveys such as DESI and Euclid, for which the volumes
of individual data chunks will be 
comparable to the entire 
volume of BOSS DR12, c.f.~\cite{Chudaykin:2019ock,Aghamousa:2016zmz}.~\footnote{Notably, the DESI constraints are primarily informed by LRG galaxies, whose properties are very similar to those of the CMASS and LOWZ samples,
so it is natural to extrapolate our results
to DESI. 
In contrast, the ELG galaxies, one of the main targets of the Euclid mission, have a somewhat different morphology, so an accurate assessment of biases for ELGs requires a dedicated analysis.
}
As we show here, in this situation  
the prior volume effects should not 
significantly complicate the interpretation of the 
results from these surveys.

\section{Comparison of different analyses choices and priors}
\label{sec:comp}

We start off by specifying 
the relations between different EFT parameters in the WC and EC analysis pipelines.

We start with bias parameters. 
The WC and EC groups utilize different sets of galaxy bias parameters introduced in~\cite{Senatore:2014eva} and \cite{Mirbabayi:2014zca}.
The WC basis of galaxy biases $(b_1^{WC},b_2^{WC},b_3^{WC},b_4^{WC})$ can be related to the EC basis $(b_1^{EC},b_2^{EC},b^{EC}_{\mathcal{G}_2},b^{EC}_{\Gamma_3})$ as follows
\be
\begin{split}
& b_1^{WC}=b_1^{EC}\,,\quad b_2^{WC} =b_1^{EC} + \frac{7}{2} b^{EC}_{\mathcal{G}_2}\,,\\
& b_3^{WC}=b_1^{EC} + 15b^{EC}_{\mathcal{G}_2} + 6b^{EC}_{\Gamma_3}\,,\\
& b_4^{WC} = \frac{1}{2}b^{EC}_{2} - \frac{7}{2} b^{EC}_{\mathcal{G}_2}\,.
\end{split} 
\ee
In what follows we will refer to the linear bias in both models simply as $b_1$. 
We re-express the nonlinear biases in the EC model,
\be
\begin{split}
 & b^{EC}_{2} =-2(b_1-b_2^{WC}-b_4^{WC})\,, \\
 & b^{EC}_{\mathcal{G}_2}  =-\frac{2}{7}(b_1-b_2^{WC})\,, \\
 & b^{EC}_{\Gamma_3} = \frac{1}{42 }(23 b_1-30 b_2^{WC}+7 b_3^{WC})\,.
\end{split}
\ee

The WC model assumes the following priors on bias parameters~\cite{DAmico:2019fhj}  
\be
\begin{split}
& \sqrt{2}(b_2^{WC}+b_4^{WC})\sim [-4,4]_{\rm flat}\,, \\
&\sqrt{2}(b_2^{WC}-b_4^{WC})\sim \N(0,2^2)\,, \\
& b_3^{WC} \sim  \N(0,2^2)~\,.
\end{split}
\ee
where $\N(m,\sigma^2)$ is a Gaussian prior centered on $m$ with a standard deviation $\sigma$, and the subscript `flat' implies a flat prior with the stated boundaries.
These can be roughly translated into 
\be
\label{ECbias}
\begin{split}
& b^{EC}_{2}\sim [-5.7,5.7]_{\rm flat}\,,\\
& b^{EC}_{\mathcal{G}_2} \in \frac{1}{7}\frac{1}{\sqrt{2}}\left[\N(0,2^2)+[-4,4]_{\rm flat}\right]\approx [-0.4,0.4]_{\rm flat}\,,\\
& b^{EC}_{\Gamma_3}\in  \approx [-1,1]_{\rm flat}\,.
\end{split} 
\ee
We approximate these boundaries by gaussian priors, 
which will be discussed shortly. 

We now proceed to the description of the counterterms and stochasticity parameters.
The exact relations for those coefficients mildly depend on redshift.
In what follows we will focus on the $z_3$ redshift bin ($z_\eff=0.61$).
For the $z_1$ redshift slice ($z_\eff=0.38$) we found the similar results which we will report in the end of the section. 
In our calculations we adopt the fiducial $\Lambda$CDM cosmology used in the official BOSS analysis~\cite{BOSS:2016wmc}: $\Omega_m=0.31$, $h=0.676$, $\omega_b=0.022$, $\sigma_8=0.8$ and $n_s=0.97$. We also adopt $b_1=2.2$ from the same BOSS analysis.

The monopole counterterm in different models is specified as follows
\be 
\begin{split}
& P_{0}^{\text{ctr}.}\Big|_{\rm EC}=-c_0k^2 \times 2\cdot \int_{0}^{1} d\mu~P_{11}(k,z) \\
& P_{0}^{\text{ctr}.}\Big|_{\rm WC}=c_{ct}\frac{k^2}{k_M^2}\times 2\cdot \int_{0}^{1} Z_1(\mu)d\mu~P_{11}(k,z)\,,
\end{split}
\ee
where~\cite{DAmico:2019fhj} 
assumed $k_M=0.7\hMpc$
and $P_{11}$ stands for the linear matter power spectrum at redshift $z$. In what follows we will suppress the explicit 
redshift dependence of $P_{11}$.
Matching 
them, we find the following relation between $c_0$ and $c_{ct}$ parameters,
\be
\frac{c_0}{[\Mpc/h]^2} = - c_{ct} \frac{1}{k_{M}^2}b_1\left(1+\frac{1}{3}\frac{f}{b_1}\right)\approx -5 c_{ct}
\ee
The counterterm for the quadrupole moment reads
\be 
\begin{split}
& P_{2}^{\text{ctr}.}\Big|_{\rm EC}=-c_2k^2 \times 2\cdot 5f\int_{0}^{1} d\mu~\mu^2L_2(\mu)P_{11}(k) \\
& P_{2}^{\text{ctr}.}\Big|_{\rm WC}=c_{r,1}\frac{k^2}{k_M^2}\times 2\cdot 5\int_{0}^{1} Z_1(\mu)d\mu~\mu^2L_2(\mu)P_{11}(k)\,,
\end{split}
\ee
from which we find 
\be
\frac{c_2}{[\Mpc/h]^2}  = - \frac{c_{r,1}}{f} \frac{1}{k_{M}^2}b_1\left(1+\frac{6}{7}\frac{f}{b_1}\right)\approx -7.4\times c_{r,1}\,.
\ee

As far as the stochastic power spectrum is concerned, both models incorporate a constant shot noise contribution and a scale-dependent redshift-space stochastic counterterm. 
The latter contributes to the quadrupole moment as follows
\be
\begin{split}
& P_{2}^{\text{stoch}.}\Big|_{\rm EC}=a_2\frac{k^2}{k_{\rm NL}^2}\frac{2}{3}
\frac{1}{\bar n_{\rm EC}} \\
& P_{2}^{\text{stoch}.}\Big|_{\rm WC}=c_{\epsilon,\,\text{quad}}\frac{k^2}{k_{M}^2} f \frac{1}{\bar n_{\rm WC}} 
\end{split}
\ee
where $k_{\rm NL}=0.45\hMpc$, $\bar n_{\rm WC}$ and $\bar n_{\rm EC}$ refer the mean galaxy densities in the WC and EC models. 
Notice that WC assume $\bar n_{\rm WC}^{-1}=2500\,(h^{-1}\Mpc)^3$ and $2200\,(h^{-1}\Mpc)^3$ for the LOWZ and CMASS redshift bins, whereas in the EC pipeline $\bar n_{\rm EC}^{-1}=3500\,(h^{-1}\Mpc)^3$ and $5000\,(h^{-1}\Mpc)^3$ for $z_1$ and  $z_3$ slices.
Note that the actual measurements 
from data give $\bar n^{-1}=4800$
and $3500$ $(h^{-1}\Mpc)^3$ for CMASS and LOWZ, respectively~\cite{Chen:2024vuf}. 
The values of $\bar n$
for the $z_1/z_3$ data split
in the EC pipeline
correspond to the mean
geometric number densities, 
i.e. the total number of galaxies 
divided by the effective geometric volume. These numbers 
are in good agreement 
with $\bar n(z_{\rm eff})$  measured 
directly from data
using the appropriate weights
and selection functions. 

Using the above values, for $z_3$ redshift bin we find the following relation
\be
a_2^{z3}=  \frac{3}{2}c_{\epsilon,\,\text{quad}} f\frac{k^2_{\rm NL}}{k_{M}^2} \frac{\bar n_{\rm EC}} {\bar n_{\rm WC}} \approx 0.22 c_{\epsilon,\,\text{quad}}\,.
\ee
For the $z_1$ chunk we get
\be
a_2^{z1}\approx 0.32 c_{\epsilon,\,\text{quad}}\,.
\ee
Adopting $c_{\epsilon,\,\text{quad}}\sim\N(0,2^2)$ from~\cite{DAmico:2019fhj}, we observe 
that the WC prior on $a_2$ is tighter than that in the EC model which is $a_2\sim\mathcal{N}(0,1^2)$. 
The inconsistency of this choice with 
the physics of the HOD-based galaxies
was discussed in~\cite{Ivanov:2024xgb}.
As we will see later, this tight prior reduces the parameter shifts caused by marginalization effects in a multi-chunk full-shape EFT analysis.
Importantly, the WC setup also excludes a scale-dependent stochastic counterterm in the monopole, $c_{\epsilon,\,\text{mono}}=0$. Although the real-space scale-dependent noise term can be set to zero 
based on the simulation results~\cite{Schmittfull:2018yuk,Schmittfull:2020trd,Ivanov:2024hgq,Ivanov:2024xgb}, the contribution from the $k^2\mu^2$ counterterm
is actually non-negligible,
which raises concerns about the 
choice $c_{\epsilon,\,\text{mono}}=0$.
Indeed, from the mathematical point of view, it is inconsistent 
to neglect a $k^2$ stochastic 
counterterm in the monopole
if the $k^2$ stochastic 
contribution is present in the quadrupole,
because physically the have the same 
origin in the $(k,\mu)$
space. 

Note that the EC pipeline features the next-to-leading order $k^4$ redshift-space counterterm $\tilde c$ and the real space scale-dependent stochastic noise counterterm $a_0$. These are defined as:
\begin{align}
& P^{\rm ctr}_{\nabla^4_{\hat{\bf z}}\delta}(k,\mu)=
-\tilde{c}k^4\mu^4 f^4(b_1+f\mu^2)^2 P_{11}(k)\,,\\
& P^{\rm stoch}(k,\mu)=\frac{1}{\bar n}\left(1+P_{\rm shot}+a_0\frac{k^2}{k_S^2}+a_2\mu^2\frac{k^2}{k_S^2}\right)\,,
\end{align}
with $k_S=0.45~\hMpc$.
The addition of these parameters has minimal impact on the inference of cosmological parameters in the base $\L$CDM model.
Nonetheless, the EC pipeline incorporates both $\tilde c$ and $a_0$ contributions for two main reasons: (a) to obtain unbiased inference of the EFT nuisance parameters and (b) to robustly estimate the parameter uncertainties.

The importance
of the $k^4\mu^4 P_{11}$
counterterm for consistency 
has been discussed
in~\cite{Ivanov:2019pdj,Chudaykin:2020aoj,Chudaykin:2020hbf,Taule:2023izt,Ivanov:2024xgb}.
First, it is used to address the strong finger-of-the-God effect, which breaks the naive EFT counting and thus necessitates the inclusion of the next-to-leading order contribution.
This term effectively captures the two-loop contribution induced by higher-derivative terms in the non-linear RSD mapping. 
As a result, its inclusion extends the validity of perturbation theory and reduces sensitivity to high-$k$ modes affected by the two-loop contributions. 
Second, dropping the $k^4\mu^4 P_{11}$ term biases the inference of the EFT parameters, in particular $c_{\epsilon,~\text{quad}}~(a_2)$,
and $c_{r,1}~(c_2)$
with which it is partly degenerate. 
While the $k^4\mu^4 P_{11}$ and $k^2\mu^2$ terms are degenerate at the level of the monopole and quadrupole spectra, this degeneracy is broken by including the hexadecapole which allows one to measure both these terms independently. 
Thus, the $k^4\mu^4 P_{11}$ term is important for the unbiased inference of both the leading-order counterterm coefficients and the $k^2\mu^2$ stochastic contribution.

To understand what drives the difference in the results between the WC and EC pipelines we perform a series of analyses with different choice of the priors and scale cuts.
Below we specify the analysis choices explored in this work.

\paragraph{WC model:} 
We adopt the exact priors form the 
~\cite{DAmico:2019fhj,Colas:2019ret}, with an analysis of $P_0$ and $P_2$. 
Below we provide the WC priors on the EFT parameters for each BOSS chunk across the $z_1/z_3$ redshift bins (if the same prior is applied in both redshifts, we omit the redshift identifier),
\be
\label{WCprior}
\begin{split}
& \frac{c_0}{[\Mpc/h]^2}\sim \N(0,10^2)\,, \frac{c_2}{[\Mpc/h]^2}\sim \N(0,60^2)\,, \\
&a_2^{z_1}\sim \N(0,0.63^2)\,, a_2^{z_3}\sim \N(0,0.44^2)\\
& P_\sh^{z_1}\sim \N(0,0.11^2)\,,  P_{\sh}^{z_3}\sim \N(0,0.08^2)\,,\\
& b_1\sim[0,4]_{\rm flat}\,, b_2\sim \N(0,5^2)\,, b_{\mathcal{G}_2}\sim \N(0,0.4^2)\,, \\
&b_{\Gamma_3}\sim \N\left(\frac{23}{42}(b_1-1),1^2\right)\,,\\
\end{split}
\ee
where we approximated the boundaries in eq.~\eqref{ECbias} by Gaussian priors.
Note that the stochastic contributions $P_{\shot}$ and $a_2$ are expressed in terms of the inverse mean galaxy density of the BOSS survey within the EC model.~\footnote{The WC team uses a significantly narrower prior on $P_\shot$ with a standard deviation of $400\,(h^{-1}\Mpc)^3$~\cite{DAmico:2019fhj}. Surprisingly, this does not introduce a significant bias in the inference of cosmological parameters
in the power-spectrum analyses of the $\Lambda$CDM model. The bias could be significant
in extensions of this setup.}
In total, the WC model has 8 EFT parameters per sky chunk when fitting two multipoles.~\footnote{The most recent WC pipeline~\cite{DAmico:2022osl} incorporates correlations between EFT parameters from different 
survey patches and redshifts, 
which help mitigate projection effects.
In this work, we compare the WC model from the original analyses~\cite{DAmico:2019fhj,Colas:2019ret} and treat all priors on the EFT parameters as uncorrelated.}

Let us now specify the scale cuts in the WC model.
The original analyses~\cite{DAmico:2019fhj,Colas:2019ret} exploit a slightly different separation into LOWZ ($z_\eff=0.32$) and CMASS ($z_\eff=0.57$) redshift slices, and specify $\kmax^{\rm LOWZ}=0.2\hMpc$ and $\kmax^{\rm CMASS}=0.23\hMpc$, respectively. Since our $z_3$ slice has effectively a slightly higher redshift and a less data volume than CMASS, we re-scale the original $\kmax^{\rm CMASS}=0.23\hMpc$ to $\kmax^{\rm z_3}=0.25\hMpc$ using the rescaling procedure suggested in Eq. (40) by~\cite{DAmico:2019fhj}
for $z=0.61$.
This rescaling 
has been used in many analyses. For instance, it 
has been employed in~\cite{Simon:2022lde}.
For that reason, 
we focus on $\kmax^{\rm z_3}=0.23\hMpc$
in the main text.
The analyses using a more conservative scale cut $\kmax^{\rm z_3}=0.23\hMpc$ are presented in App.~\ref{sec:kmax}.
There we show that 
practically there
is no difference between
$\kmax^{\rm z_3}=0.23\hMpc$
and $\kmax^{\rm z_3}=0.25\hMpc$.
We keep $\kmax^{\rm z_1}=0.2\hMpc$ following~\cite{Simon:2022lde}.

\paragraph{EC1 model:} 
We use the WC priors but consistently include a
scale-dependent stochastic counterterm in the monopole. 
Specifically, we add
\be
P_{0}^{\text{stoch}.}\Big|_{\rm EC}=a_2\frac{k^2}{k_{\rm NL}^2}\frac{1}{3}
\frac{1}{\bar n_{\rm EC}} \\
\ee
where $a_2$ is taken from~\eqref{WCprior}.
In the WC notation, it corresponds to $c_{\epsilon,\,\text{mono}}\sim\N(0,1^2)$.
The other aspects of the analysis, including the data cuts, remain the same as in the WC model.

\paragraph{EC2 model:} 
We adopt the EC priors for the galaxy power spectrum from \cite{Philcox:2021kcw},
\be
\begin{split}
& \frac{c_0}{[\Mpc/h]^2}\sim \N(0,30^2)\,, \frac{c_2}{[\Mpc/h]^2}\sim \N(30,30^2)\,, \\
&\frac{\tilde c}{[\Mpc/h]^4}\sim \N(0,500^2)\,,P_\sh\sim \N(0,1^2)\,,\\
& a_0\sim \N(0,1^2)\,, a_2\sim \N(0,1^2)\,,\\
& b_1\sim[0,4]_{\rm flat}\,,  b_2\sim \N(0,1^2)\,, b_{\mathcal{G}_2}\sim \N(0,1^2)\,,\\ 
& b_{\Gamma_3}\sim \N\left(\frac{23}{42}(b_1-1),1^2\right)\,,\\
\end{split}
\ee
In total, the EC2 priors consist of 10 parameters.
We use the exact same scale cuts and number of multipoles as in the WC and EC1 models.

\paragraph{EC3 model:} 
We implement the baseline EC analysis pipeline \cite{Philcox:2021kcw} which additionally includes the power spectrum hexadecapole~\cite{Chudaykin:2020ghx}, 
the 
real-space power spectrum proxy~\cite{Ivanov:2021haa}, and the bispectrum monopole~\cite{Ivanov:2021kcd,Philcox:2021kcw}. 
We adopt the EC2 priors for the galaxy power spectrum defined above.
We introduce four extra EFT parameters associated with the additional probes, and impose the following priors,
\be
\begin{split}
& \frac{c_4}{[\Mpc/h]^2}\sim \N(0,30^2)\,, \frac{c_1}{[\Mpc/h]^2}\sim \N(0,5^2)\,,\\
& A_\sh\sim \N(0,1^2)\,, B_\sh\sim \N(0,1^2)\,.\\
\end{split}
\ee
In total, the EC3 priors consist of 14 parameters.

Following the EC pipeline \cite{Philcox:2021kcw}, we change the scale cuts for the galaxy power spectrum to be $\kmax=0.2\hMpc$ for the $z_1/z_3$ redshift bins.
As we will see later, using a lower value of $\kmax^{z_3}$ effectively reduces the systematic bias in parameter recovery.
We also adopt $\kmax=0.08\hMpc$ for the bispectrum measurements.

Note that for any realistic analysis we 
recommend using the complete analysis pipeline presented in \cite{Philcox:2021kcw}, including the 
full data vector. This corresponds to the EC3 model. 
The removal of 
datasets from the total datavector used in 
\cite{Philcox:2021kcw},
e.g. the bispectrum, results 
in
enhanced 
parameter degeneracies,
which increase the bias 
in parameter recovery.

\section{Tests on PT Challenge data}
\label{sec:ptc}

In this section we test the analysis pipelines using the PT Challenge simulation data. 
The PT Challenge simulation suite was designed to reproduce clustering properties 
of the BOSS-like galaxies from the DR12 sample~\cite{Nishimichi:2020tvu}. The total simulation volume is $V=566\,(\Gpch)^3$, which is about 100 times larger than the cumulative volume of the galaxy BOSS DR12 sample $V=5.8\,(\Gpch)^3$. 
The large simulation volume provides a stringent test of the theoretical model, allowing us to separate the theory systematic error from the statistical error.
This helps assess the origin of parameter shifts and accurately quantify the impact of priors in various analysis setups.
We analyze the mean power spectrum and bispectrum over 10 random realizations,
for two redshift redshifts, $z=0.38$ and $z=0.61$ that match the 
effective redshift bins $z_1$ and $z_3$ of BOSS DR12.
Since the $k$-binning in the challenge power spectra is quite wide ($\Delta k=0.01\,\Mpch$) compared to the fundamental mode of the simulation box, we average our theoretical predictions over each bin as in~\cite{Nishimichi:2020tvu,Philcox:2021kcw}.  
 
We perform two types of analysis. 
First, we present an analysis of the PT Challenge simulation using the analytical Gaussian covariance.\footnote{We neglect the cross-correlation between the power spectrum and bispectrum in the covariance which was shown to be a good approximation for the scales of interest~\cite{Ivanov:2021kcd}.} We adopt the volume and number density specific to the PT Challenge simulation, and adjust the priors accordingly. Second, we carry out a multi-chunk analysis of the four simulation samples using the actual BOSS data covariance estimated from the Patchy mocks.
This setup aims to simulate the analysis of the actual BOSS data, allowing us to estimate the impact of priors in the analysis of the real data.

\subsection{PT Challenge analysis}
\label{sec:ptc1}

In this analysis we adjust the priors by using the actual value of the mean number density in the PT challenge simulation, $\bar n_{\rm PTC}^{-1}=2300\,(h^{-1}\Mpc)^3$.~\footnote{The mean galaxy densities
are slightly different at $z_1$ and $z_3$, but this difference is insignificant for the purposes of our analysis.}
We also use this value to build up the analytical Gaussian covariance for both the power spectrum and bispectrum.

As a first step, we fit the PT challenge data taken from a single snapshot at $z_3$ ($z_\eff=0.61$) with the different analysis pipelines.
We perform the analysis using the Gaussian covariance which corresponds to the cumulative volume of the BOSS survey, $5.8\,(\Gpch)^3$, and to the total simulation volume, $566\,(\Gpch)^3$.

\begin{figure*}
    \centering
    \includegraphics[width=0.99\textwidth]{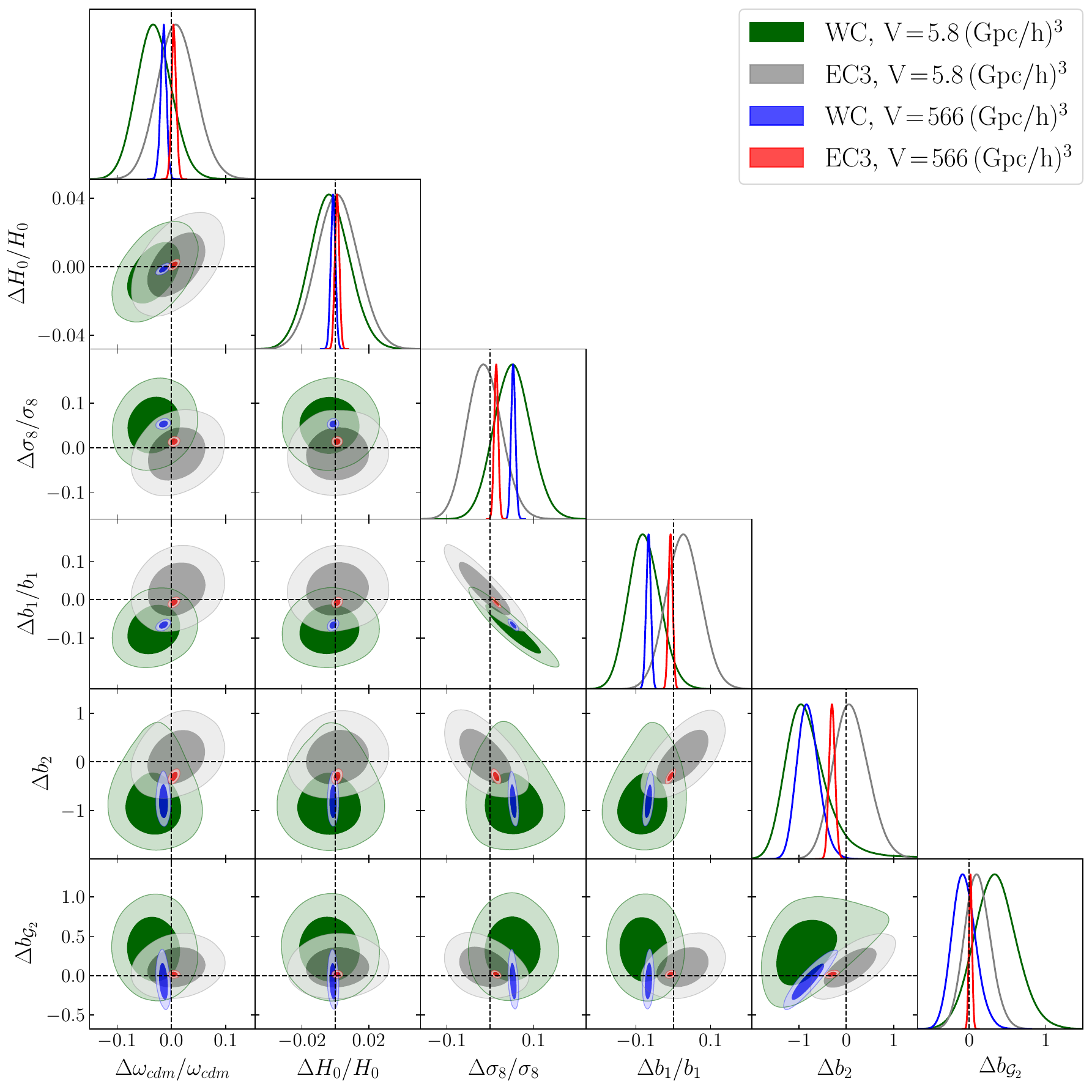}
    \caption{
    Posterior distributions inferred from the single snapshot of the PT challenge simulation at $z_\eff=0.61$ in the WC and EC3 models using the analytical covariance matching the BOSS survey volume 5.8~($h^{-1}$Gpc)$^3$, and the 
    covariance corresponding to the total simulation volume $566\,(\Gpch)^3$.   
    }
    \label{fig:PTCgauss}
\end{figure*}

Our results in the WC and EC3 models are shown in Fig.~\ref{fig:PTCgauss}.
The 1d marginalized constraints on cosmological and bias parameters are given in Tab.~\ref{tab:PTCgauss}.
\begin{table*}[t!]
%First part -- WC, EC1
\begin{tabular}{|c||c c c| c c c |}
    \hline
    \multirow{3}{*}{\diagbox[width=2.5cm]{\small Param.}{\small Model}}  
    & \multicolumn{3}{c|}{\multirow{2}{*}{WC}}
    & \multicolumn{3}{c|}{\multirow{2}{*}{EC1}}
    \\ 
    & & &
    & & &
    \\ %[0.2cm]
    \cline{2-7}
    & $4\!\times\! V=5.8$ & $V=5.8$ & $V=566$
    & $4\!\times\! V=5.8$ & $V=5.8$ & $V=566$
    \\ \hline
$\Delta\omega_{cdm}/\omega_{cdm}$  
& $-0.038_{-0.033}^{+0.030}$ % WCcor 4V
& $-0.031_{-0.034}^{+0.032}$ % WCcor
& $-0.015_{-0.0052}^{+0.0058}$ % WCcor
& $-0.028_{-0.035}^{+0.031}$ % EC1cor 4V
& $-0.015_{-0.037}^{+0.034}$ % 
& $-0.011_{-0.0050}^{+0.0045}$ %
\\ 
\hline 
$\Delta h/h$  
& $-0.004_{-0.012}^{+0.012}$ % WCcor 4V
& $-0.003_{-0.013}^{+0.012}$ % WCcor
& $-0.0014_{-0.0014}^{+0.0015}$ % WCcor 
& $-0.002_{-0.013}^{+0.012}$ % EC1cor 4V
& $-0.001_{-0.013}^{+0.012}$ %
& $-0.0032_{-0.0014}^{+0.0014}$ %
\\ \hline
$\Delta \sigma_8/\sigma_8$   
& $0.029_{-0.042}^{+0.040}$ % WCcor 4V
& $0.051_{-0.044}^{+0.041}$ % WCcor 
& $0.053_{-0.0051}^{+0.0049}$ % WCcor 
& $0.027_{-0.043}^{+0.040}$ % EC1cor 4V
& $0.058_{-0.044}^{+0.041}$ %
& $0.062_{-0.0046}^{+0.0045}$ %
\\ \hline\hline
$\Delta b_1/b_1$   
& $-$
& $-0.077_{-0.046}^{+0.041}$ % WCcor
& $-0.066_{-0.0062}^{+0.0065}$ % WCcor
& $-$
& $-0.069_{-0.046}^{+0.042}$ % 
& $-0.078_{-0.0049}^{+0.0050}$ %
\\ \hline
$\Delta b_2$   
& $-$
& $-0.77_{-0.61}^{+0.31}$ %% WCcor
& $-0.80_{-0.25}^{+0.21}$ %% WCcor
& $-$
& $-0.38_{-0.62}^{+0.37}$ %
& $-0.31_{-0.18}^{+0.17}$ %
\\ \hline
$\Delta b_{\mathcal{G}_2}$  
& $-$
& $0.34_{-0.29}^{+0.27}$ %% WCcor
& $-0.07_{-0.17}^{+0.15}$ %% WCcor
& $-$
& $0.49_{-0.28}^{+0.27}$ %
& $0.92_{-0.14}^{+0.14}$ %
\\ 
\hline\hline
\end{tabular}\par\vskip-1.4pt
%Second part -- EC2, EC3
\begin{tabular}{|c||c c c| c| c c c|}
    \hline
    \multirow{3}{*}{\diagbox[width=2.5cm]{\small Param.}{\small Model}} 
    & \multicolumn{3}{c|}{\multirow{2}{*}{EC2}}
    & EC2
    & \multicolumn{3}{c|}{\multirow{2}{*}{EC3}}
    \\ 
    & & &
    & $k_{\max}=0.2$
    & & &
    \\ %[0.2cm]
    \cline{2-8}
    & $4\!\times\! V=5.8$ & $V=5.8$ & $V=566$
    & $V=566$
    & $4\!\times\! V=5.8$ & $V=5.8$ & $V=566$
    \\ \hline
$\Delta\omega_{cdm}/\omega_{cdm}$   
& $-0.001_{-0.042}^{+0.039}$ % EC2 4V
& $-0.006_{-0.048}^{+0.044}$ %
& $-0.0085_{-0.0070}^{+0.0064}$ %
& $-0.0066_{-0.0078}^{+0.0076}$ %
& $0.002_{-0.034}^{+0.031}$ % EC3 4V
& $0.010_{-0.037}^{+0.035}$ %
& $0.0048_{-0.0046}^{+0.0043}$ % 
\\ 
\hline 
$\Delta h/h$   
& $0.001_{-0.013}^{+0.013}$ % % EC2 4V
& $0.0_{-0.014}^{+0.013}$ %
& $0.0002_{-0.0017}^{+0.0017}$ %
& $-0.0034_{-0.0019}^{+0.0019}$ %
& $0.001_{-0.013}^{+0.012}$ % EC3 4V
& $0.001_{-0.013}^{+0.012}$ %
& $0.0010_{-0.0013}^{+0.0013}$ %
\\ \hline
$\Delta \sigma_8/\sigma_8$   
& $-0.018_{-0.046}^{+0.042}$ % % EC2 4V
& $0.037_{-0.047}^{+0.044}$ %
& $0.051_{-0.0053}^{+0.0051}$ % 
& $0.026_{-0.0060}^{+0.0057}$ %
& $-0.053_{-0.035}^{+0.031}$ % EC3 4V
& $-0.013_{-0.042}^{+0.039}$ %
& $0.014_{-0.0048}^{+0.0047}$ %
\\ \hline\hline
$\Delta b_1/b_1$  
& $-$
& $-0.044_{-0.051}^{+0.049}$ %
& $-0.057_{-0.0089}^{+0.0086}$ % 
& $-0.038_{-0.0087}^{+0.0091}$ %
& $-$
& $0.027_{-0.048}^{+0.046}$ %
& $-0.0075_{-0.0059}^{+0.0057}$ %
\\ \hline
$\Delta b_2$   
& $-$
& $-0.33_{-0.80}^{+0.66}$ %
& $-0.81_{-0.31}^{+0.29}$ % 
& $-0.62_{-0.43}^{+0.36}$ %
& $-$
& $0.11_{-0.41}^{+0.35}$ %%
& $-0.30_{-0.066}^{+0.064}$ %% 
\\ \hline
$\Delta b_{\mathcal{G}_2}$   
& $-$
& $0.25_{-0.37}^{+0.37}$ % 
& $-0.29_{-0.26}^{+0.26}$ % 
& $0.43_{-0.32}^{+0.32}$ %
& $-$
& $0.11_{-0.18}^{+0.17}$ %%
& $0.016_{-0.021}^{+0.020}$ %% 
\\ 
\hline
\end{tabular}
\caption{Marginalized 1d constraints on the cosmological and bias parameters inferred from the single- and multi-chunk analyses of the PT Challenge simulation data for different models.
The cosmological parameters are shown as $(p - p_{\rm fid.}) / p_{\rm fid.}$, 
where $p_{\rm fid.}$ denotes the fiducial value used in the simulation. 
For single-chunk analyses we also report constraints on the bias parameters.
All volumes are expressed in units of $\,[\Gpch]^3$.
}
\label{tab:PTCgauss}
\end{table*}

Let us first discuss the analysis with the full simulation volume.
We observe that the WC and EC3 pipelines yield significantly different 
posteriors. In particular, one can notice a significant, 5\% bias 
in the recovery of $\sigma_8$ when the WC model is used.
In the EC3 model, the systematic error associated with $\sigma_8$ is significantly smaller, being around $1\%$.
The measurement of $\omega_{\rm cdm}$ in the WC model shows a bias of 1.5\%, whereas, in the EC3 pipeline, this shift is reduced to 0.5\%. 
The recovery of $h$ is unbiased in both WC and EC3 models, even when considering the total simulation volume.
The second important observation
is that the recovery of nuisance parameters is also biased when using the WC model.~\footnote{We consider estimates
of nuisance parameters 
from the galaxy-matter cross power spectrum
and the real space bispectrum~\cite{Ivanov:2021kcd,Philcox:2022frc} as their ``ground truth'' values.} 
Namely, $b_2$ is lower than ground truth by $\sim 1$, and  $b_1$ is shifted toward lower values, a shift correlated with the bias in $\sigma_8$.
The EC3 pipeline demonstrates much better agreement with all values of bias parameters estimated in previous bispectrum analyses~\cite{Ivanov:2021kcd,Philcox:2022frc,Ivanov:2023qzb}.

In the analysis with a covariance matrix rescaled to match the cumulative volume of the BOSS survey,
we again observe significant differences in the posteriors between the WC and EC3 models.
Specifically, in the WC model, $\omega_{cdm}$ and $\sigma_8$ deviate from their true values by $\sigma_{\rm BOSS}$ 
and $1.2\sigma_{\rm BOSS}$, respectively, in terms of the statistical error from the analysis with the BOSS volume.
Notably, the mean value of $\omega_{cdm}$ is lower compared to the results from the full simulation volume.
As we will demonstrate later, this shift arises from the fact that the WC model enforces $c_{\epsilon,\,\text{mono}}=0$, and this bias is fully corrected after including a scale-dependent stochastic counterterm in the monopole. 
The bias in $\sigma_8$ recovery is entirely driven by the theory systematic error on this parameter.
No prior volume effect on $\sigma_8$ is observed for the WC model.
The situation is quite different for the EC3 model, where the shifts in cosmological parameters do not exceed $0.3\sigma_{\rm BOSS}$.
Interestingly, the projection effects for $\sigma_8$ work in the opposite direction, reducing the small systematic bias present in the $V=566$ case.
Unlike the WC model, the EC3 priors have a negligible impact on $\omega_{\rm cdm}$ due to its consistent treatment of stochastic biases.

The recovery of nuisance parameters remains 
biased for the WC model, even for the errorbars corresponding to the volume of the BOSS survey. In particular, the $b_{\mathcal{G}_2}-b_2$ contour is shifted away 
from the ground truth by 2$\sigma$. These large shifts are related to the shifts in 
$\omega_{\rm cdm}$ and $\sigma_8$ (for the WC model) and could be especially 
problematic for the analyses of non-local primordial non-Gaussianity (PNG), 
which is correlated with $b_{\mathcal{G}_2}$ and $b_2$~\cite{Cabass:2022wjy}.
In contrast, the EC3 model allows for the recovery of all bias parameters within $\lesssim0.5\sigma_{\rm BOSS}$.
In fact, the PNG analyses 
based on the \texttt{CLASS-PT} code
~\cite{Cabass:2022wjy,Cabass:2022ymb}
used a somewhat lower power 
spectrum scale cut $\kmax=0.17~\hMpc$,
which completely eliminates the bias 
on the EFT parameters, c.f.~\cite{Philcox:2022frc}.

The magnitude of projection effects increases with the number of nuisance parameters, leading the single-chunk analysis to somewhat underestimate the parameter shifts due to marginalization effects.  
To reliable assess the prior volume effects, we perform a multi-chunk analysis of four simulation samples -- two at the $z_1$ redshift bin ($z_\eff=0.38$) and two at the $z_3$ redshift bin ($z_\eff=0.61$).
For each data chunk, we introduce an independent set of EFT parameters, allowing them to vary freely in the analysis.
The mock catalogues at the same redshift slice are designed to be identical emulating the BOSS samples in the Northern and Southern galactic caps (NGC and SGC).
We adjust the volume for each chunk to match the actual BOSS volumes of the NGCz1, SGCz1, NGCz3, and SGCz3 caps, while keeping the PT Challenge mean galaxy density $\bar n_{\rm PTC}$ to allow for meaningful comparisons with previous results. We call this analysis $4\times V=5.8$.

\begin{figure*}
    \centering
    \includegraphics[width=0.48\textwidth]{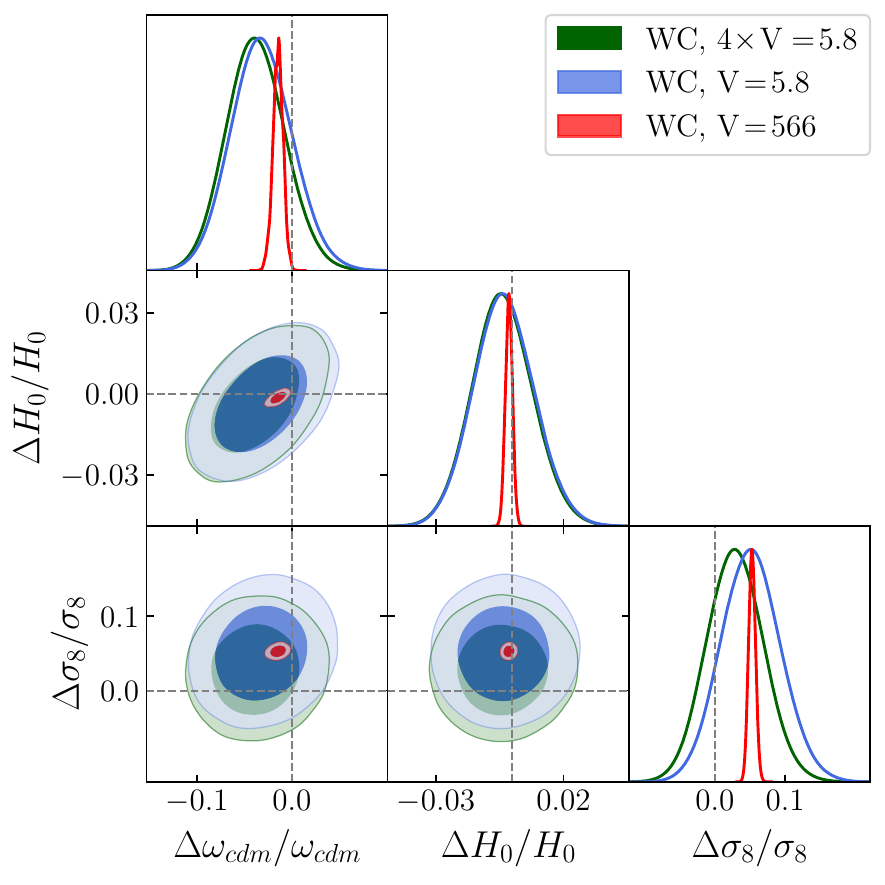}
    \includegraphics[width=0.48\textwidth]{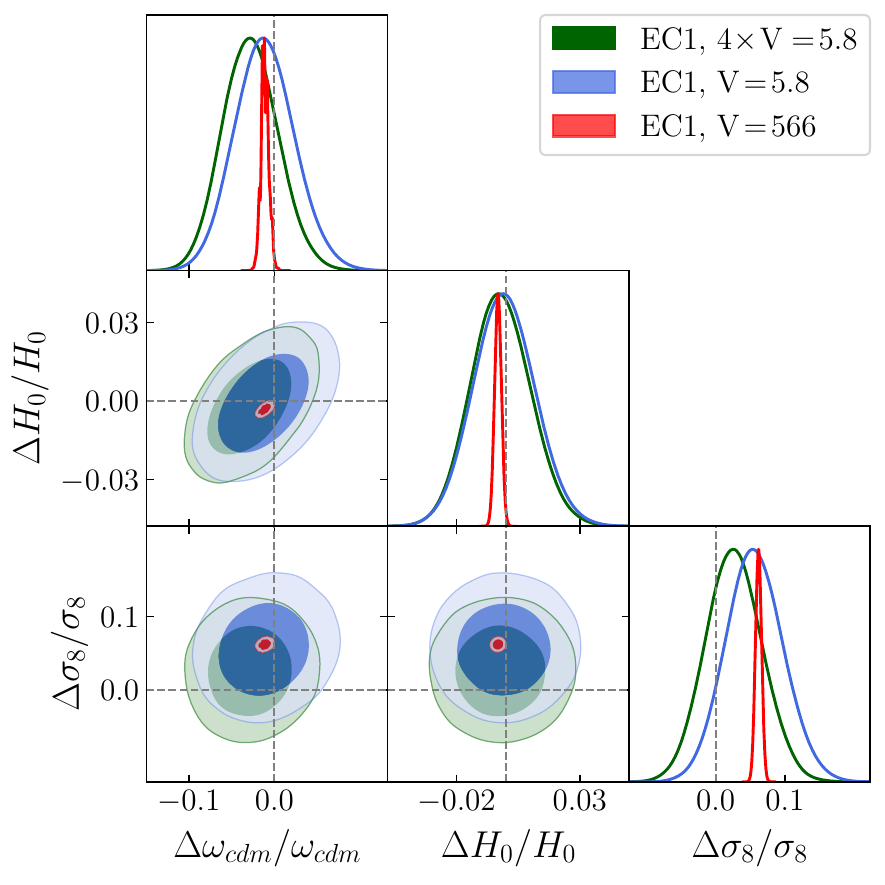}
    \includegraphics[width=0.48\textwidth]{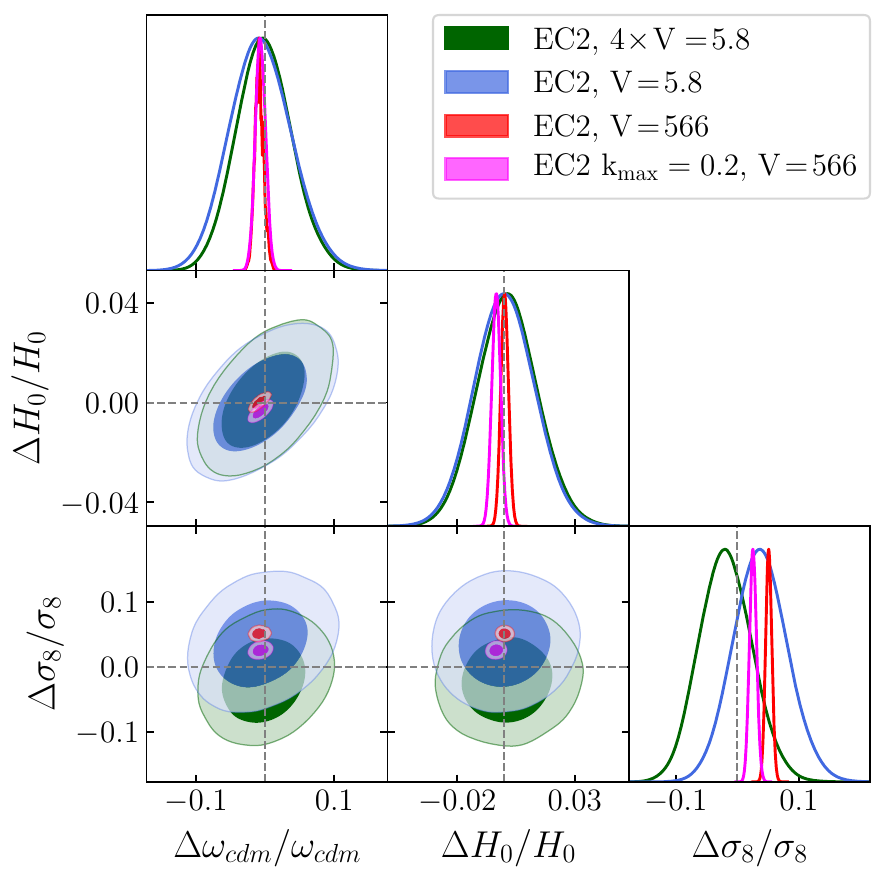}
    \includegraphics[width=0.48\textwidth]{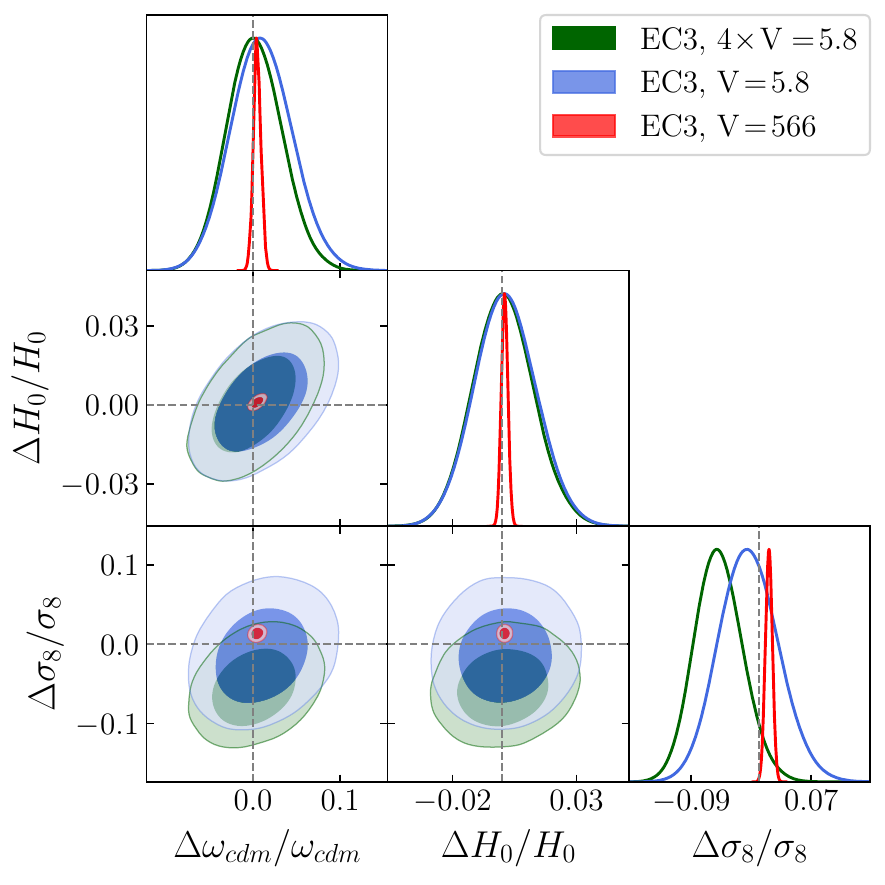}
    \caption{Posterior distributions from the different analyses of the PT challenge simulation for the WC model ({\it upper left panel}), the EC1 model ({\it upper right panel}), the EC2 model ({\it lower left panel}) and the EC3 model ({\it lower right panel}). The results are inferred from the four BOSS-like data chunks (green), as well as from the single snapshot at $z_\eff=0.61$ with the  covariance matching the BOSS survey volume (blue), and the covariance rescaled to match the full PT challenge volume 566~($h^{-1}$Gpc)$^3$ (red).
    For the EC2 model, we also show the analysis with $\kmax=0.2\hMpc$ and the covariance corresponding to the full simulation volume (in pink).
    }
    \label{fig:PTCmulti}
\end{figure*}

The results of the multi-chunk analysis for the different models are presented in Fig.~\ref{fig:PTCmulti}.  
The corresponding 1d marginalized parameter constraints are given in Tab.~\ref{tab:PTCgauss}.

Let us quantify the projection effects for each model.
We assess the amplitude of projection effects by calculating the difference in the parameter means between the $4\times V$ and $V=566$ analyses, and expressing the result in units of the standard deviation from the $4\times V$ analysis.~\footnote{In the $V=566$ analysis, the difference between the mean and best-fit values of parameters is negligible.} 
Starting with the WC model, we observe that 
the mean of $\omega_{cdm}$ in the multi-chunk analysis is shifted lower compared to the $V=566$ case.
However, this shift does not scale with the number of boxes (or equivalently, with the number of nuisance parameters), therefore it cannot be explained by the usual marginalization effects caused by non-Gaussian correlations in the posteriors. 
Additionally, we identify a $0.6\sigma$ downward shift in $\sigma_8$ due to projection effects.
Unlike the WC model, the EC1 setup shows unbiased recovery of $\omega_{\rm cdm}$ in the $V=5.8$ analysis.
This suggests that adding a scale-dependent stochastic counterterm in the monopole is essential for unbiased inference of cosmological parameters.
At the same time, allowing more flexibility in the stochasticity model slightly increases the Bayesian marginalization shift in $\omega_{\rm cdm}$ and $\sigma_8$ to $0.5\sigma$ and $0.8\sigma$, respectively.

The EC2 models shows a $1.6\sigma$ shift in $\sigma_8$ due to projection effects.
Notably, the actual theory systematic bias in the recovery of $\sigma_8$ persists at the same 5\% level observed in the WC model, indicating that this shift cannot be explained by the choice of priors.
To elucidate the source of this bias, we perform an additional analysis using EC2 priors but with $\kmax=0.2\hMpc$ for both $z_1/z_3$ redshift bins, similar to the EC3 setup.
This reduces the systematic bias in $\sigma_8$ by half, bringing it down to $2.6\%$.
Thus, the bias on $\sigma_8$ is largely driven by the overoptimistic 
scale cuts in the WC model.
In the EC3 model, we detect the projection effects for $\sigma_8$ at the $2.2\sigma$ level, while the other parameters remain unaffected. 
However, the mean of $\sigma_8$ is shifted by only $1.7\sigma$ from the ground truth, an acceptable shift in a typical Bayesian analysis.
Importantly, the theory-driven systematic error on $\sigma_8$ is reduced to $1.4\%$, compared to $2.6\%$ in the EC2 $k_{\max}=0.2$ analysis.
This improvement is attributed to the inclusion of bispectrum data, which 
breaks the degeneracies present at the power spectrum level and 
pulls the parameters closer to their true values.

All in all, we conclude that the WC model is somewhat restrictive. 
First, the posterior of $\omega_{cdm}$ is affected by the priors generated by the inaccuracy of the WC stochasticity model.
Second, the WC model applies overoptimistic scale cuts in the $z_3$ redshift bin which significantly bias parameter recovery, overestimating $\sigma_8$ by more than $5\%$.
Unlike the Bayesian marginalization effect, this is a genuine
systematic bias that does not vanish with better priors or larger data volumes.
In contrast, the EC3 model leads to a small ($<1.5\%$) systematic bias
in the recovery of cosmological parameters. This bias constitutes a small 
negligible fraction of the statistical error in the actual BOSS data.
Using more observables, like the cross-correlation with CMB lensing or  
imposing informative 
simulation-based priors on nuisance parameters from~\cite{Ivanov:2024xgb} can mitigate the Bayesian marginalization effects present in the EC3 analysis with the BOSS volume.

\subsection{BOSS-like analysis}
\label{sec:ptc2}

Here, we explore the implications of Bayesian marginalization effects and theory systematic bias within a realistic analysis framework.  
To that end, we perform a multi-chunk analysis of the four simulation samples using the actual BOSS data covariance estimated from the Patchy mocks.
This allows us to closely reproduce the actual BOSS data analysis.
Additionally, to evaluate the systematic error, we perform a separate analysis using Gaussian covariance corresponding to 100 times the actual volume of each BOSS-like data chunk, yielding a cumulative volume of $580\,(\Gpch)^3$. 
For this analysis, we adopt the actual galaxy number density of the PT Challenge simulation to reduce the impact of marginalization projection effects.
Note that we do not utilize the empirical covariance from the Patchy mocks in the $100\times V_{\rm BOSS}$ analysis due to sample noise, which is particularly significant for bispectrum measurements.
We refer to these analyses as $V_{\rm BOSS}$ and $100\times V_{\rm BOSS}$, respectively.

\begin{figure*}[t!]
    \centering
    \includegraphics[width=0.6\textwidth]{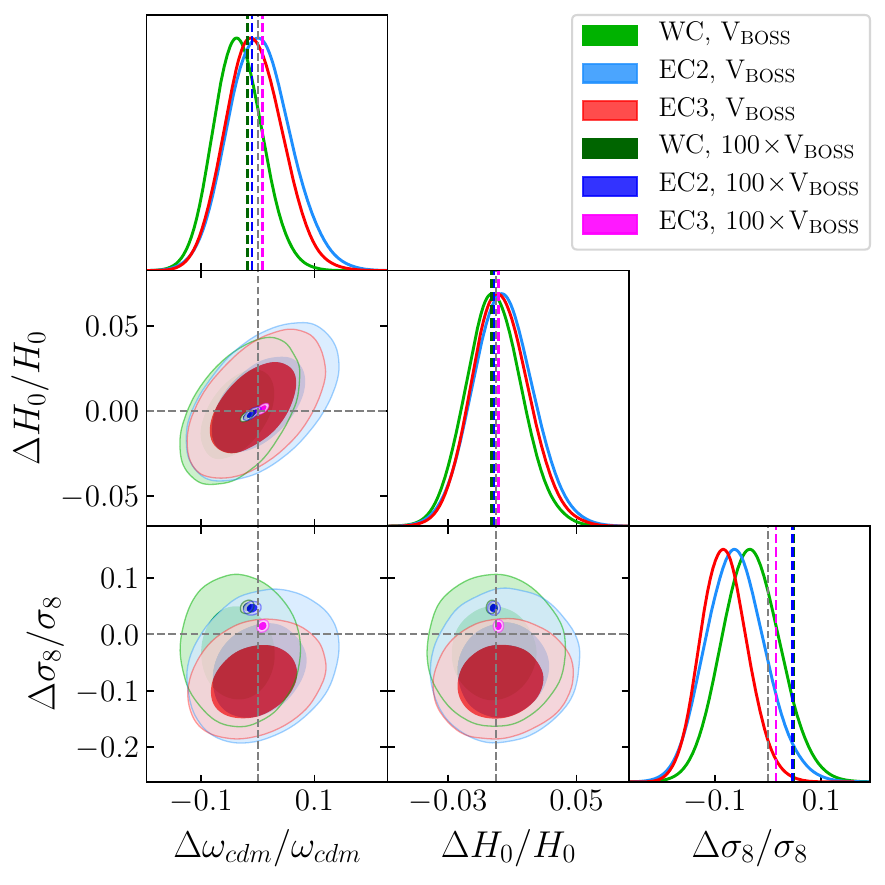}
    \caption{
    Posterior distributions inferred from the four BOSS-like samples of the PT Challenge simulation using the BOSS data covariance estimated from the Patchy mocks ($V_{\rm BOSS}$) and the analytical Gaussian covariance corresponding to 100 times the BOSS volume ($100\times V_{\rm BOSS}$). 
    For $100\times V_{\rm BOSS}$, only the mean values of the corresponding parameters are depicted by dashed lines.   
    }
    \label{fig:ptc_all}
\end{figure*}

Our results are shown in Fig.~\ref{fig:ptc_all}.
The marginalized parameter constraints are listed in Tab.~\ref{tab:PTC}.
\begin{table*}[t!]
  \begin{tabular}{|c||c c|c c|c c|} \hline
   \multirow{2}{*}{\diagbox{ {\small Param.}}{\small Model}}  
   & \multicolumn{2}{c}{WC}
   & \multicolumn{2}{|c|}{EC2} 
   & \multicolumn{2}{c|}{EC3}
 \\ [0.2cm]
 \cline{2-7}
 & $V_{\rm BOSS}$
 & $100\times V_{\rm BOSS}$
 & $V_{\rm BOSS}$
 & $100\times V_{\rm BOSS}$
 & $V_{\rm BOSS}$
 & $100\times V_{\rm BOSS}$
 \\
\hline
$\Delta\omega_{cdm}/\omega_{cdm}$   
& $-0.034_{-0.047}^{+0.041}$ %% WC covBOSS
& $-0.018_{-0.005}^{+0.005}$ %% WC 100xVboss
& $0.004_{-0.060}^{+0.052}$ %% EC2 covBOSS
& $-0.0098_{-0.0065}^{+0.0061}$ %% EC2 100xVboss
& $-0.006_{-0.054}^{+0.048}$ %% EC3 covBOSS
& $0.0088_{-0.0047}^{+0.0043}$ %% EC3 100xVboss
\\ 
\hline 
$\Delta h/h$   
& $-0.001_{-0.018}^{+0.017}$ %% WC covBOSS
& $-0.0028_{-0.0014}^{+0.0014}$ %% WC 100xVboss
& $0.004_{-0.020}^{+0.019}$ %% EC2 covBOSS
& $-0.0015_{-0.0017}^{+0.0016}$ %% EC2 100xVboss
& $0.003_{-0.019}^{+0.017}$ %% EC3 covBOSS
& $0.0014_{-0.0014}^{+0.0014}$ %% EC3 100xVboss 
\\ \hline
$\Delta \sigma_8/\sigma_8$   
& $-0.031_{-0.058}^{+0.054}$ %% WC covBOSS
& $0.049_{-0.005}^{+0.005}$ %% WC 100xVboss
& $-0.060_{-0.061}^{+0.053}$ %% EC2 covBOSS
& $0.046_{-0.005}^{+0.005}$ %% EC2 100xVboss
& $-0.082_{-0.047}^{+0.041}$ %% EC3 covBOSS
& $0.015_{-0.005}^{+0.005}$ %% EC3 100xVboss
\\ 
\hline
\end{tabular}
\caption{Marginalized 1d constraints on the cosmological parameters from the BOSS-like analyses of the four simulation samples for three models and two choices of the covariance matrix.
All parameters are shown as $(p - p_{\rm fid.}) / p_{\rm fid.}$, where $p_{\rm fid.}$ represents the fiducial value used in the simulations.
}
\label{tab:PTC}
\end{table*}
In the WC model, we observe a $0.8\sigma$ shift in $\omega_{cdm}$ from the ground truth for the covariance of the $V_\mathrm{BOSS}$ volume, caused by the inaccuracy of the WC template.
We identify a $5\%$ systematic error on $\sigma_8$ from the $100\times V_\mathrm{BOSS}$ analysis, in perfect agreement with the results from the PT Challenge analysis.
In contrast, the EC2 model provides an unbiased measurement of $\omega_{cdm}$ due to the consistent treatment of stochastic bias.
The EC3 model leads to a small systematic bias in the recovery of cosmological parameters, with the largest $1.5\%$ error on $\sigma_8$.
However, the broader priors in EC3 induce a somewhat larger marginalization projection shift of about $2\sigma$ for $\sigma_8$.
This shift is generally acceptable in a Bayesian analysis. It can be reduced by imposing correlated priors on the EFT parameters, as discussed in~\cite{DAmico:2022osl}.
An alternative and more motivated from the Bayesian 
point of view approach is to employ simulation-based priors from~\cite{Ivanov:2024xgb,Ivanov:2024xgb}.

Let us briefly summarize the previous tests of the WC analysis pipeline performed in the literature.
As a first step, the Refs.~\cite{DAmico:2019fhj,DAmico:2022osl} validated the analysis pipeline using the Patchy Mocks~\cite{Kitaura:2015uqa}, which did not show any significant bias in parameter recovery. 
This is expected, as the Patchy Mocks use an approximate gravity solver and thus can not reliably predict clustering on mildly non-linear scales.~\footnote{The Patchy Mocks underestimate the power of the power spectrum monopole at $k\sim0.2\hMpc$ compared to that measured from the BOSS DR12 data~\cite{Kitaura:2015uqa}.}
Secondly, the Refs.~\cite{DAmico:2022osl} analyzed the mean of 84 Nseries mocks~\cite{BOSS:2016wmc}.
These mocks are based on full N-body simulations and implement the effects of the survey selection function and survey geometry.
They reported a $4.7\%$ bias
in the $\sigma_8$
recovery 
and a $1.3\%$ lower bias
in the $\omega_{cdm}$
recovery 
from the power
spectrum using their pipeline in complete 
agreement with our 
findings. 
That said, the interpretations of the results are different. 
Ref.~\cite{DAmico:2022osl} 
quantifies the parameter shifts in terms of the statistical uncertainty from the analysis of the CMASS NGC data chunk,
while here we compare these shifts 
to the statistical error from the actual BOSS data analysis (including
all four BOSS data chunks). Clearly, 
the significance of the 
bias in parameter recovery
when expressed in units of the standard deviation 
is larger in our case than in the analysis of~\cite{DAmico:2022osl}.

Let us compare now our results with the relatively recent analysis of projection effects performed in~\cite{Simon:2022lde}. 
The key difference with our approach is that their analysis uses synthetic data generated from the best-fit model to BOSS data.
\cite{Simon:2022lde} report a downward shift in the mean value of $\sigma_8$ at the $1.2\sigma$ level for the WC model and at the $1.8\sigma$ level for the EC model.
Our results, presented in Tab.~\ref{tab:PTC}, show similar marginalization shifts: $1.4\sigma$ for the WC model and $2.2\sigma$ for the EC3 model.~\footnote{Ref.~\cite{Simon:2022lde} estimates the projections effects by calculating the parameter differences w.r.t. the best-fit prediction. To provide a meaningful comparison with their results, we compute the marginalization shift here as the difference in parameter means between the $V_{\rm BOSS}$ and $100\times V_{\rm BOSS}$ analyses. We have checked that for $100\times V_{\rm BOSS}$ the difference between the mean and best-fit values of parameters is negligible.} 
Given the differences in our analysis methodologies (e.g., Ref.~\cite{Simon:2022lde} does not use bispectrum measurements), this is a good agreement.
The virtue of the present analysis is that it is based on simulation data, allowing us to directly estimate the theory-driven systematic bias in parameter recovery.

Let us also point out that Ref.~\cite{Simon:2022lde} did not include the bispectrum
and did not consistently use the $\kmax=0.2~\hMpc$
in their reproduction
of the analysis pipeline based on the \texttt{CLASS-PT} code. 
Both of these choices lead to a larger best-fit value of $\sigma_8$,
which agrees with the WC results, but is actually biased higher than the true value by $\approx 5\%$.
The comparison of EC and WC analysis choices
in these inconsistent settings 
may lead to misleading conclusions. 
This issue is particularly important in the context of profile likelihood analysis, where a recent study of early dark energy~ \cite{Herold:2022iib} used  $\kmax=0.25\hMpc$, which, as we show here, introduces a significant bias in 
the $\sigma_8$ recovery.

\section{Conclusion}
\label{sec:concl}

In this work, we explore the different analysis choices associated with the \texttt{CLASS-PT} and \texttt{PyBird} codes. 
Analysis settings fall into two main categories: EFT modeling (EFT parameters and priors) and data cuts. Below, we outline the implications of these settings for a realistic BOSS-like analysis.

The EFT modeling of the two fitting pipelines differs in several aspects. First, the EC model includes two additional EFT parameters: the next-to-leading order $k^4$ counterterm $\tilde c$ and the real space scale-dependent stochastic noise counterterm $a_0$.
Second, the priors on EFT parameters vary significantly between the EC and WC pipelines, with the WC model employing tighter priors than those used in the EC model.
However, we find that these differences have only a marginal impact on the resulting parameter constraints, which validates the prior choices made in both analyses.
A final and more important difference is that the WC model neglects the scale-dependent stochastic contribution to the power spectrum monopole, which leads to a $0.8\sigma$ downward shift in $\omega_{cdm}$ in the realistic BOSS-like analysis. 
In contrast, the EC model consistently incorporates scale-dependent stochastic bias and yields an unbiased inference of $\omega_{cdm}$. 
Thus, it is important to use a consistent model for stochasticity, which requires including a $k^2$ stochastic 
counterterm in both the monopole and quadrupole.

Another important aspect of EFT-based full-shape analyses is the choice of the momentum-scale cut $\kmax$.
Using overoptimistic scale cuts amplifies theory-driven systematic error, which can significantly bias cosmological constraints.
The WC model adopts $\kmax=0.25\hMpc$ in the $z_3$ redshift bin, which significantly biases parameter recovery, overestimating $\sigma_8$ by $\approx5\%$ (equivalent to $1\sigma$), or by $\approx4\%$ when using the $\kmax=0.23\hMpc$ from Refs.~\cite{DAmico:2019fhj,Colas:2019ret}.
By contrast, the EC team employs a more conservative data cut, $\kmax=0.20\hMpc$, for the same data chunk, which results in only a small ($<1.5\%$) systematic bias in the recovery of cosmological parameters. 
This occurs because the theoretical error associated with the two-loop corrections grows rapidly with $\kmax$, so that even a modest increase of $\kmax$ from $0.2~\hMpc$ to $0.25~\hMpc$ (or $0.23~\hMpc$) significantly amplifies the error, making it a significant source
of systematic uncertainty.~\footnote{Using a larger value of $\kmax$ in the WC analysis alleviates prior volume projection effects, but at the cost of increasing the theory-driven systematic bias in parameter recovery.}
We conclude that the difference in scale cuts is the main source of the mismatch between the EC and WC analyses, a point that has been largely overlooked in the literature.
This discrepancy in $\kmax$ choices also produces differences in non-Gaussianity constraints between the EC~\cite{Cabass:2022wjy,Cabass:2022ymb} and WC~\cite{DAmico:2022gki} analyses, which has been later confirmed in studies of the one-loop bispectrum~\cite{Bakx:2025pop} and PNG inference from DESI data~\cite{Chudaykin:2025vdh}.
We conclude that conservative scale cuts are essential to achieve unbiased inference of cosmological parameters.

\textit{Acknowledgments.}
MI thanks George Efstathiou for inspiring discussions.
We thank Guido D'Amico for useful discussions. 
We are additionally grateful to the anonymous referee for useful comments, which helped improve the presentation.
AC acknowledges funding from the Swiss National Science Foundation.
All numerical calculations have been performed with the Helios cluster at the Institute for Advanced Study, Princeton.
This work was supported in part by MEXT/JSPS KAKENHI Grant
Numbers JP20H05861, JP21H01081,
JP22K03634, JP24H00215 and JP24H00221.

\appendix 

\section{Analysis with $\kmax^{\rm z_3}=0.23\hMpc$}
\label{sec:kmax}

We perform the BOSS-like analyses in the WC and EC2 models using a more conservative scale cut $\kmax^{\rm z_3}=0.23\hMpc$ from the original analyses~\cite{DAmico:2019fhj,Colas:2019ret}.

We carried out a multi-chunk analysis of the four simulation samples using the BOSS data covariance estimated from the Patchy mocks. 
Additionally, we perform the analysis with the Gaussian covariance corresponding to 100 times the actual volume of each BOSS-like data chunks.
We adopt $\kmax^{\rm z_1}=0.2\hMpc$ for the $z_1$ redshift bin, as in the main analysis, but use a more conservative scale cut $\kmax^{\rm z_3}=0.23\hMpc$ in the $z_3$ redshift bin.
Further details of the analysis is given in Sec.~\ref{sec:ptc2}.

\begin{figure*}[t!]
    \centering
    \includegraphics[width=0.6\textwidth]{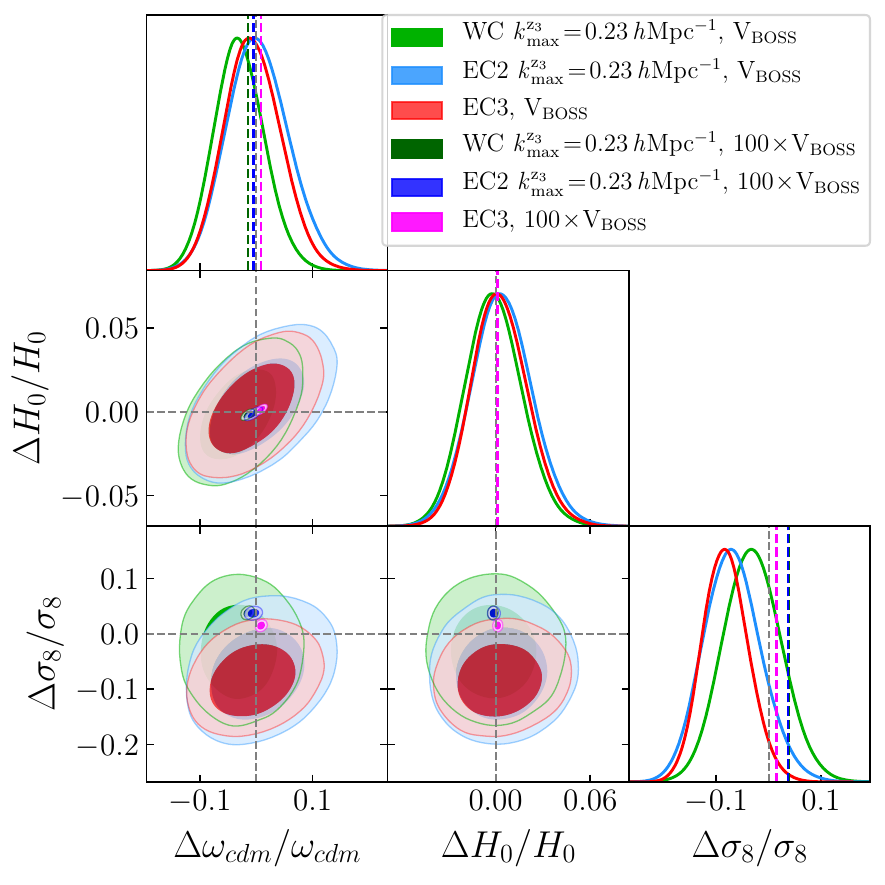}
    \caption{
    Posterior distributions inferred from the four BOSS-like samples of the PT Challenge simulation using the BOSS data covariance estimated from the Patchy mocks ($V_{\rm BOSS}$) and the analytical Gaussian covariance corresponding to 100 times the BOSS volume ($100\times V_{\rm BOSS}$). 
    For $100\times V_{\rm BOSS}$, only the mean values of the corresponding parameters are depicted by dashed lines.  
    For the WC and EC2 models, the $\kmax^{\rm z_1}/\kmax^{\rm z_3}=0.2/0.23\hMpc$ are adopted, while for the EC3 model $\kmax=0.2\hMpc$ is used for both redshift bins. 
    }
    \label{fig:ptc_all2}
\end{figure*}

Our results in the WC and EC2 models are presented in Fig.~\ref{fig:ptc_all2}.
We also show the posteriors for the EC3 model which uses $\kmax=0.2\hMpc$ for both $z_1/z_3$ redshift bins.
The marginalized parameter constraints are listed in Tab.~\ref{tab:PTC2}.
\begin{table*}[t!]
  \begin{tabular}{|c||c c|c c|} \hline
   \multirow{2}{*}{\diagbox{ {\small Param.}}{\small Model}}  
   & \multicolumn{2}{c}{\begin{tabular}{c}
    WC \\
    $\kmax^{\rm z_3}=0.23\hMpc$
  \end{tabular}}
   & \multicolumn{2}{|c|}{\begin{tabular}{c}
    EC2 \\
    $\kmax^{\rm z_3}=0.23\hMpc$
  \end{tabular}} 
   % & \multicolumn{2}{c|}{EC3}
 \\ [0.2cm]
% &
 \cline{2-5}
 & $V_{\rm BOSS}$
 & $100\times V_{\rm BOSS}$
 & $V_{\rm BOSS}$
 & $100\times V_{\rm BOSS}$
 % & $V_{\rm BOSS}$
 % & $100\times V_{\rm BOSS}$
 \\
\hline
$\Delta\omega_{cdm}/\omega_{cdm}$   
& $-0.029_{-0.048}^{+0.044}$ %% WC covBOSS kmax023
& $-0.015_{-0.005}^{+0.005}$ %% WC 100xVboss kmax023
& $0.004_{-0.061}^{+0.053}$ %% EC2 covBOSS kmax023
& $-0.0048_{-0.0067}^{+0.0069}$ %% EC2 100xVboss kmax023
% & $-0.006_{-0.054}^{+0.048}$ %% EC3 covBOSS
% & $0.0088_{-0.0047}^{+0.0043}$ %% EC3 100xVboss
\\ 
\hline 
$\Delta h/h$   
& $-0.001_{-0.019}^{+0.018}$ %% WC covBOSS kmax023
& $-0.0017_{-0.0014}^{+0.0014}$ %% WC 100xVboss kmax023
& $0.004_{-0.020}^{+0.019}$ %% EC2 covBOSS kmax023
& $-0.0011_{-0.0017}^{+0.0017}$ %% EC2 100xVboss kmax023
% & $0.003_{-0.019}^{+0.017}$ %% EC3 covBOSS
% & $0.0014_{-0.0014}^{+0.0014}$ %% EC3 100xVboss 
\\ \hline
$\Delta \sigma_8/\sigma_8$   
& $-0.030_{-0.059}^{+0.055}$ %% WC covBOSS kmax023
& $0.038_{-0.005}^{+0.005}$ %% WC 100xVboss kmax023
& $-0.069_{-0.060}^{+0.053}$ %% EC2 covBOSS kmax023
& $0.037_{-0.005}^{+0.005}$ %% EC2 100xVboss kmax023
% & $-0.082_{-0.047}^{+0.041}$ %% EC3 covBOSS
% & $0.015_{-0.005}^{+0.005}$ %% EC3 100xVboss
\\ 
\hline
\end{tabular}
\caption{Marginalized 1d constraints on the cosmological parameters from the BOSS-like analyses of the four simulation samples for the WC/EC2 models and two choices of the covariance matrix.
All parameters are shown as $(p - p_{\rm fid.}) / p_{\rm fid.}$, where $p_{\rm fid.}$ represents the fiducial value used in the simulations.
}
\label{tab:PTC2}
\end{table*}

We found that the parameter constraints obtained with the scale cuts from Refs.~\cite{DAmico:2019fhj,Colas:2019ret} are fully consistent with those from the baseline analysis employing $\kmax^{\rm z_3}=0.25\hMpc$ , cf. Tab.~\ref{tab:PTC}.
We identify a $4\%$ theory systematic bias in $\sigma_8$, which is slightly smaller than the $5\%$ bias observed in the main analysis.
Thus, adopting $\kmax^{\rm z_3}=0.23\hMpc$ does not illuminate the bias in the $\sigma_8$ recovery. 
We therefore conclude that the scale cuts $\kmax^{\rm z_3}=0.23$ and $0.25\hMpc$ are overoptimistic and result in a significant systematic error in the estimation of $\sigma_8$. As shown in Sec.~\ref{sec:ptc1}, this bias can be reduced by half when using $\kmax=0.2\hMpc$ for both $z_1/z_3$ redshift bins in the EC2 model.

\bibliography{short.bib}

\begin{thebibliography}{70}%
\makeatletter
\providecommand \@ifxundefined [1]{%
 \@ifx{#1\undefined}
}%
\providecommand \@ifnum [1]{%
 \ifnum #1\expandafter \@firstoftwo
 \else \expandafter \@secondoftwo
 \fi
}%
\providecommand \@ifx [1]{%
 \ifx #1\expandafter \@firstoftwo
 \else \expandafter \@secondoftwo
 \fi
}%
\providecommand \natexlab [1]{#1}%
\providecommand \enquote  [1]{``#1''}%
\providecommand \bibnamefont  [1]{#1}%
\providecommand \bibfnamefont [1]{#1}%
\providecommand \citenamefont [1]{#1}%
\providecommand \href@noop [0]{\@secondoftwo}%
\providecommand \href [0]{\begingroup \@sanitize@url \@href}%
\providecommand \@href[1]{\@@startlink{#1}\@@href}%
\providecommand \@@href[1]{\endgroup#1\@@endlink}%
\providecommand \@sanitize@url [0]{\catcode `\\12\catcode `\$12\catcode
  `\&12\catcode `\#12\catcode `\^12\catcode `\_12\catcode `\%12\relax}%
\providecommand \@@startlink[1]{}%
\providecommand \@@endlink[0]{}%
\providecommand \url  [0]{\begingroup\@sanitize@url \@url }%
\providecommand \@url [1]{\endgroup\@href {#1}{\urlprefix }}%
\providecommand \urlprefix  [0]{URL }%
\providecommand \Eprint [0]{\href }%
\providecommand \doibase [0]{http://dx.doi.org/}%
\providecommand \selectlanguage [0]{\@gobble}%
\providecommand \bibinfo  [0]{\@secondoftwo}%
\providecommand \bibfield  [0]{\@secondoftwo}%
\providecommand \translation [1]{[#1]}%
\providecommand \BibitemOpen [0]{}%
\providecommand \bibitemStop [0]{}%
\providecommand \bibitemNoStop [0]{.\EOS\space}%
\providecommand \EOS [0]{\spacefactor3000\relax}%
\providecommand \BibitemShut  [1]{\csname bibitem#1\endcsname}%
\let\auto@bib@innerbib\@empty
%</preamble>
\bibitem [{\citenamefont {Ivanov}\ \emph
  {et~al.}(2020{\natexlab{a}})\citenamefont {Ivanov}, \citenamefont
  {Simonovi\'c},\ and\ \citenamefont {Zaldarriaga}}]{Ivanov:2019pdj}%
  \BibitemOpen
  \bibfield  {author} {\bibinfo {author} {\bibfnamefont {M.~M.}\ \bibnamefont
  {Ivanov}}, \bibinfo {author} {\bibfnamefont {M.}~\bibnamefont {Simonovi\'c}},
  \ and\ \bibinfo {author} {\bibfnamefont {M.}~\bibnamefont {Zaldarriaga}},\
  }\href {\doibase 10.1088/1475-7516/2020/05/042} {\bibfield  {journal}
  {\bibinfo  {journal} {JCAP}\ }\textbf {\bibinfo {volume} {05}},\ \bibinfo
  {pages} {042} (\bibinfo {year} {2020}{\natexlab{a}})},\ \Eprint
  {http://arxiv.org/abs/1909.05277} {arXiv:1909.05277 [astro-ph.CO]}
  \BibitemShut {NoStop}%
\bibitem [{\citenamefont {D'Amico}\ \emph {et~al.}(2019)\citenamefont
  {D'Amico}, \citenamefont {Gleyzes}, \citenamefont {Kokron}, \citenamefont
  {Markovic}, \citenamefont {Senatore}, \citenamefont {Zhang}, \citenamefont
  {Beutler},\ and\ \citenamefont {Gil-Marín}}]{DAmico:2019fhj}%
  \BibitemOpen
  \bibfield  {author} {\bibinfo {author} {\bibfnamefont {G.}~\bibnamefont
  {D'Amico}}, \bibinfo {author} {\bibfnamefont {J.}~\bibnamefont {Gleyzes}},
  \bibinfo {author} {\bibfnamefont {N.}~\bibnamefont {Kokron}}, \bibinfo
  {author} {\bibfnamefont {D.}~\bibnamefont {Markovic}}, \bibinfo {author}
  {\bibfnamefont {L.}~\bibnamefont {Senatore}}, \bibinfo {author}
  {\bibfnamefont {P.}~\bibnamefont {Zhang}}, \bibinfo {author} {\bibfnamefont
  {F.}~\bibnamefont {Beutler}}, \ and\ \bibinfo {author} {\bibfnamefont
  {H.}~\bibnamefont {Gil-Marín}},\ }\href@noop {} {\  (\bibinfo {year}
  {2019})},\ \Eprint {http://arxiv.org/abs/1909.05271} {arXiv:1909.05271
  [astro-ph.CO]} \BibitemShut {NoStop}%
%%CITATION = ARXIV:1909.05271;%%
\bibitem [{\citenamefont {Philcox}\ \emph {et~al.}(2020)\citenamefont
  {Philcox}, \citenamefont {Ivanov}, \citenamefont {Simonovi\'c},\ and\
  \citenamefont {Zaldarriaga}}]{Philcox:2020vvt}%
  \BibitemOpen
  \bibfield  {author} {\bibinfo {author} {\bibfnamefont {O.~H.~E.}\
  \bibnamefont {Philcox}}, \bibinfo {author} {\bibfnamefont {M.~M.}\
  \bibnamefont {Ivanov}}, \bibinfo {author} {\bibfnamefont {M.}~\bibnamefont
  {Simonovi\'c}}, \ and\ \bibinfo {author} {\bibfnamefont {M.}~\bibnamefont
  {Zaldarriaga}},\ }\href {\doibase 10.1088/1475-7516/2020/05/032} {\bibfield
  {journal} {\bibinfo  {journal} {JCAP}\ }\textbf {\bibinfo {volume} {05}},\
  \bibinfo {pages} {032} (\bibinfo {year} {2020})},\ \Eprint
  {http://arxiv.org/abs/2002.04035} {arXiv:2002.04035 [astro-ph.CO]}
  \BibitemShut {NoStop}%
\bibitem [{\citenamefont {Chen}\ \emph {et~al.}(2022)\citenamefont {Chen},
  \citenamefont {Vlah},\ and\ \citenamefont {White}}]{Chen:2021wdi}%
  \BibitemOpen
  \bibfield  {author} {\bibinfo {author} {\bibfnamefont {S.-F.}\ \bibnamefont
  {Chen}}, \bibinfo {author} {\bibfnamefont {Z.}~\bibnamefont {Vlah}}, \ and\
  \bibinfo {author} {\bibfnamefont {M.}~\bibnamefont {White}},\ }\href
  {\doibase 10.1088/1475-7516/2022/02/008} {\bibfield  {journal} {\bibinfo
  {journal} {JCAP}\ }\textbf {\bibinfo {volume} {02}},\ \bibinfo {pages} {008}
  (\bibinfo {year} {2022})},\ \Eprint {http://arxiv.org/abs/2110.05530}
  {arXiv:2110.05530 [astro-ph.CO]} \BibitemShut {NoStop}%
\bibitem [{\citenamefont {Philcox}\ and\ \citenamefont
  {Ivanov}(2022)}]{Philcox:2021kcw}%
  \BibitemOpen
  \bibfield  {author} {\bibinfo {author} {\bibfnamefont {O.~H.~E.}\
  \bibnamefont {Philcox}}\ and\ \bibinfo {author} {\bibfnamefont {M.~M.}\
  \bibnamefont {Ivanov}},\ }\href {\doibase 10.1103/PhysRevD.105.043517}
  {\bibfield  {journal} {\bibinfo  {journal} {Phys. Rev. D}\ }\textbf {\bibinfo
  {volume} {105}},\ \bibinfo {pages} {043517} (\bibinfo {year} {2022})},\
  \Eprint {http://arxiv.org/abs/2112.04515} {arXiv:2112.04515 [astro-ph.CO]}
  \BibitemShut {NoStop}%
\bibitem [{\citenamefont {Chudaykin}\ and\ \citenamefont
  {Ivanov}(2023)}]{Chudaykin:2022nru}%
  \BibitemOpen
  \bibfield  {author} {\bibinfo {author} {\bibfnamefont {A.}~\bibnamefont
  {Chudaykin}}\ and\ \bibinfo {author} {\bibfnamefont {M.~M.}\ \bibnamefont
  {Ivanov}},\ }\href {\doibase 10.1103/PhysRevD.107.043518} {\bibfield
  {journal} {\bibinfo  {journal} {Phys. Rev. D}\ }\textbf {\bibinfo {volume}
  {107}},\ \bibinfo {pages} {043518} (\bibinfo {year} {2023})},\ \Eprint
  {http://arxiv.org/abs/2210.17044} {arXiv:2210.17044 [astro-ph.CO]}
  \BibitemShut {NoStop}%
\bibitem [{\citenamefont {Baumann}\ \emph {et~al.}(2012)\citenamefont
  {Baumann}, \citenamefont {Nicolis}, \citenamefont {Senatore},\ and\
  \citenamefont {Zaldarriaga}}]{Baumann:2010tm}%
  \BibitemOpen
  \bibfield  {author} {\bibinfo {author} {\bibfnamefont {D.}~\bibnamefont
  {Baumann}}, \bibinfo {author} {\bibfnamefont {A.}~\bibnamefont {Nicolis}},
  \bibinfo {author} {\bibfnamefont {L.}~\bibnamefont {Senatore}}, \ and\
  \bibinfo {author} {\bibfnamefont {M.}~\bibnamefont {Zaldarriaga}},\ }\href
  {\doibase 10.1088/1475-7516/2012/07/051} {\bibfield  {journal} {\bibinfo
  {journal} {JCAP}\ }\textbf {\bibinfo {volume} {1207}},\ \bibinfo {pages}
  {051} (\bibinfo {year} {2012})},\ \Eprint {http://arxiv.org/abs/1004.2488}
  {arXiv:1004.2488 [astro-ph.CO]} \BibitemShut {NoStop}%
%%CITATION = ARXIV:1004.2488;%%
\bibitem [{\citenamefont {Carrasco}\ \emph {et~al.}(2012)\citenamefont
  {Carrasco}, \citenamefont {Hertzberg},\ and\ \citenamefont
  {Senatore}}]{Carrasco:2012cv}%
  \BibitemOpen
  \bibfield  {author} {\bibinfo {author} {\bibfnamefont {J.~J.~M.}\
  \bibnamefont {Carrasco}}, \bibinfo {author} {\bibfnamefont {M.~P.}\
  \bibnamefont {Hertzberg}}, \ and\ \bibinfo {author} {\bibfnamefont
  {L.}~\bibnamefont {Senatore}},\ }\href {\doibase 10.1007/JHEP09(2012)082}
  {\bibfield  {journal} {\bibinfo  {journal} {JHEP}\ }\textbf {\bibinfo
  {volume} {09}},\ \bibinfo {pages} {082} (\bibinfo {year} {2012})},\ \Eprint
  {http://arxiv.org/abs/1206.2926} {arXiv:1206.2926 [astro-ph.CO]} \BibitemShut
  {NoStop}%
%%CITATION = ARXIV:1206.2926;%%
\bibitem [{\citenamefont {Ivanov}(2023)}]{Ivanov:2022mrd}%
  \BibitemOpen
  \bibfield  {author} {\bibinfo {author} {\bibfnamefont {M.~M.}\ \bibnamefont
  {Ivanov}},\ }\enquote {\bibinfo {title} {{Effective Field Theory for
  Large-Scale Structure}},}\ \ (\bibinfo {year} {2023})\ \Eprint
  {http://arxiv.org/abs/2212.08488} {arXiv:2212.08488 [astro-ph.CO]}
  \BibitemShut {NoStop}%
\bibitem [{\citenamefont {Blas}\ \emph
  {et~al.}(2016{\natexlab{a}})\citenamefont {Blas}, \citenamefont {Garny},
  \citenamefont {Ivanov},\ and\ \citenamefont {Sibiryakov}}]{Blas:2015qsi}%
  \BibitemOpen
  \bibfield  {author} {\bibinfo {author} {\bibfnamefont {D.}~\bibnamefont
  {Blas}}, \bibinfo {author} {\bibfnamefont {M.}~\bibnamefont {Garny}},
  \bibinfo {author} {\bibfnamefont {M.~M.}\ \bibnamefont {Ivanov}}, \ and\
  \bibinfo {author} {\bibfnamefont {S.}~\bibnamefont {Sibiryakov}},\ }\href
  {\doibase 10.1088/1475-7516/2016/07/052} {\bibfield  {journal} {\bibinfo
  {journal} {JCAP}\ }\textbf {\bibinfo {volume} {1607}},\ \bibinfo {pages}
  {052} (\bibinfo {year} {2016}{\natexlab{a}})},\ \Eprint
  {http://arxiv.org/abs/1512.05807} {arXiv:1512.05807 [astro-ph.CO]}
  \BibitemShut {NoStop}%
%%CITATION = ARXIV:1512.05807;%%
\bibitem [{\citenamefont {Blas}\ \emph
  {et~al.}(2016{\natexlab{b}})\citenamefont {Blas}, \citenamefont {Garny},
  \citenamefont {Ivanov},\ and\ \citenamefont {Sibiryakov}}]{Blas:2016sfa}%
  \BibitemOpen
  \bibfield  {author} {\bibinfo {author} {\bibfnamefont {D.}~\bibnamefont
  {Blas}}, \bibinfo {author} {\bibfnamefont {M.}~\bibnamefont {Garny}},
  \bibinfo {author} {\bibfnamefont {M.~M.}\ \bibnamefont {Ivanov}}, \ and\
  \bibinfo {author} {\bibfnamefont {S.}~\bibnamefont {Sibiryakov}},\ }\href
  {\doibase 10.1088/1475-7516/2016/07/028} {\bibfield  {journal} {\bibinfo
  {journal} {JCAP}\ }\textbf {\bibinfo {volume} {1607}},\ \bibinfo {pages}
  {028} (\bibinfo {year} {2016}{\natexlab{b}})},\ \Eprint
  {http://arxiv.org/abs/1605.02149} {arXiv:1605.02149 [astro-ph.CO]}
  \BibitemShut {NoStop}%
%%CITATION = ARXIV:1605.02149;%%
\bibitem [{\citenamefont {Chen}\ \emph {et~al.}(2020)\citenamefont {Chen},
  \citenamefont {Vlah},\ and\ \citenamefont {White}}]{Chen:2020fxs}%
  \BibitemOpen
  \bibfield  {author} {\bibinfo {author} {\bibfnamefont {S.-F.}\ \bibnamefont
  {Chen}}, \bibinfo {author} {\bibfnamefont {Z.}~\bibnamefont {Vlah}}, \ and\
  \bibinfo {author} {\bibfnamefont {M.}~\bibnamefont {White}},\ }\href
  {\doibase 10.1088/1475-7516/2020/07/062} {\bibfield  {journal} {\bibinfo
  {journal} {JCAP}\ }\textbf {\bibinfo {volume} {07}},\ \bibinfo {pages} {062}
  (\bibinfo {year} {2020})},\ \Eprint {http://arxiv.org/abs/2005.00523}
  {arXiv:2005.00523 [astro-ph.CO]} \BibitemShut {NoStop}%
\bibitem [{\citenamefont {Chen}\ \emph {et~al.}(2021)\citenamefont {Chen},
  \citenamefont {Vlah}, \citenamefont {Castorina},\ and\ \citenamefont
  {White}}]{Chen:2020zjt}%
  \BibitemOpen
  \bibfield  {author} {\bibinfo {author} {\bibfnamefont {S.-F.}\ \bibnamefont
  {Chen}}, \bibinfo {author} {\bibfnamefont {Z.}~\bibnamefont {Vlah}}, \bibinfo
  {author} {\bibfnamefont {E.}~\bibnamefont {Castorina}}, \ and\ \bibinfo
  {author} {\bibfnamefont {M.}~\bibnamefont {White}},\ }\href {\doibase
  10.1088/1475-7516/2021/03/100} {\bibfield  {journal} {\bibinfo  {journal}
  {JCAP}\ }\textbf {\bibinfo {volume} {03}},\ \bibinfo {pages} {100} (\bibinfo
  {year} {2021})},\ \Eprint {http://arxiv.org/abs/2012.04636} {arXiv:2012.04636
  [astro-ph.CO]} \BibitemShut {NoStop}%
\bibitem [{\citenamefont {Wadekar}\ \emph {et~al.}(2020)\citenamefont
  {Wadekar}, \citenamefont {Ivanov},\ and\ \citenamefont
  {Scoccimarro}}]{Wadekar:2020hax}%
  \BibitemOpen
  \bibfield  {author} {\bibinfo {author} {\bibfnamefont {D.}~\bibnamefont
  {Wadekar}}, \bibinfo {author} {\bibfnamefont {M.~M.}\ \bibnamefont {Ivanov}},
  \ and\ \bibinfo {author} {\bibfnamefont {R.}~\bibnamefont {Scoccimarro}},\
  }\href {\doibase 10.1103/PhysRevD.102.123521} {\bibfield  {journal} {\bibinfo
   {journal} {Phys. Rev. D}\ }\textbf {\bibinfo {volume} {102}},\ \bibinfo
  {pages} {123521} (\bibinfo {year} {2020})},\ \Eprint
  {http://arxiv.org/abs/2009.00622} {arXiv:2009.00622 [astro-ph.CO]}
  \BibitemShut {NoStop}%
\bibitem [{\citenamefont {Cabass}\ \emph {et~al.}(2023)\citenamefont {Cabass},
  \citenamefont {Ivanov}, \citenamefont {Philcox}, \citenamefont {Simonovic},\
  and\ \citenamefont {Zaldarriaga}}]{Cabass:2022epm}%
  \BibitemOpen
  \bibfield  {author} {\bibinfo {author} {\bibfnamefont {G.}~\bibnamefont
  {Cabass}}, \bibinfo {author} {\bibfnamefont {M.~M.}\ \bibnamefont {Ivanov}},
  \bibinfo {author} {\bibfnamefont {O.~H.~E.}\ \bibnamefont {Philcox}},
  \bibinfo {author} {\bibfnamefont {M.}~\bibnamefont {Simonovic}}, \ and\
  \bibinfo {author} {\bibfnamefont {M.}~\bibnamefont {Zaldarriaga}},\ }\href
  {\doibase 10.1016/j.physletb.2023.137912} {\bibfield  {journal} {\bibinfo
  {journal} {Phys. Lett. B}\ }\textbf {\bibinfo {volume} {841}},\ \bibinfo
  {pages} {137912} (\bibinfo {year} {2023})},\ \Eprint
  {http://arxiv.org/abs/2211.14899} {arXiv:2211.14899 [astro-ph.CO]}
  \BibitemShut {NoStop}%
\bibitem [{\citenamefont {Ivanov}\ \emph
  {et~al.}(2024{\natexlab{a}})\citenamefont {Ivanov}, \citenamefont
  {Cuesta-Lazaro}, \citenamefont {Mishra-Sharma}, \citenamefont {Obuljen},\
  and\ \citenamefont {Toomey}}]{Ivanov:2024hgq}%
  \BibitemOpen
  \bibfield  {author} {\bibinfo {author} {\bibfnamefont {M.~M.}\ \bibnamefont
  {Ivanov}}, \bibinfo {author} {\bibfnamefont {C.}~\bibnamefont
  {Cuesta-Lazaro}}, \bibinfo {author} {\bibfnamefont {S.}~\bibnamefont
  {Mishra-Sharma}}, \bibinfo {author} {\bibfnamefont {A.}~\bibnamefont
  {Obuljen}}, \ and\ \bibinfo {author} {\bibfnamefont {M.~W.}\ \bibnamefont
  {Toomey}},\ }\href {\doibase 10.1103/PhysRevD.110.063538} {\bibfield
  {journal} {\bibinfo  {journal} {Phys. Rev. D}\ }\textbf {\bibinfo {volume}
  {110}},\ \bibinfo {pages} {063538} (\bibinfo {year} {2024}{\natexlab{a}})},\
  \Eprint {http://arxiv.org/abs/2402.13310} {arXiv:2402.13310 [astro-ph.CO]}
  \BibitemShut {NoStop}%
\bibitem [{\citenamefont {Cabass}\ \emph {et~al.}(2024)\citenamefont {Cabass},
  \citenamefont {Philcox}, \citenamefont {Ivanov}, \citenamefont {Akitsu},
  \citenamefont {Chen}, \citenamefont {Simonovi\'c},\ and\ \citenamefont
  {Zaldarriaga}}]{Cabass:2024wob}%
  \BibitemOpen
  \bibfield  {author} {\bibinfo {author} {\bibfnamefont {G.}~\bibnamefont
  {Cabass}}, \bibinfo {author} {\bibfnamefont {O.~H.~E.}\ \bibnamefont
  {Philcox}}, \bibinfo {author} {\bibfnamefont {M.~M.}\ \bibnamefont {Ivanov}},
  \bibinfo {author} {\bibfnamefont {K.}~\bibnamefont {Akitsu}}, \bibinfo
  {author} {\bibfnamefont {S.-F.}\ \bibnamefont {Chen}}, \bibinfo {author}
  {\bibfnamefont {M.}~\bibnamefont {Simonovi\'c}}, \ and\ \bibinfo {author}
  {\bibfnamefont {M.}~\bibnamefont {Zaldarriaga}},\ }\href@noop {} {\
  (\bibinfo {year} {2024})},\ \Eprint {http://arxiv.org/abs/2404.01894}
  {arXiv:2404.01894 [astro-ph.CO]} \BibitemShut {NoStop}%
\bibitem [{\citenamefont {Ivanov}\ \emph
  {et~al.}(2024{\natexlab{b}})\citenamefont {Ivanov}, \citenamefont {Obuljen},
  \citenamefont {Cuesta-Lazaro},\ and\ \citenamefont
  {Toomey}}]{Ivanov:2024xgb}%
  \BibitemOpen
  \bibfield  {author} {\bibinfo {author} {\bibfnamefont {M.~M.}\ \bibnamefont
  {Ivanov}}, \bibinfo {author} {\bibfnamefont {A.}~\bibnamefont {Obuljen}},
  \bibinfo {author} {\bibfnamefont {C.}~\bibnamefont {Cuesta-Lazaro}}, \ and\
  \bibinfo {author} {\bibfnamefont {M.~W.}\ \bibnamefont {Toomey}},\
  }\href@noop {} {\  (\bibinfo {year} {2024}{\natexlab{b}})},\ \Eprint
  {http://arxiv.org/abs/2409.10609} {arXiv:2409.10609 [astro-ph.CO]}
  \BibitemShut {NoStop}%
\bibitem [{\citenamefont {O'Shaughnessy}\ \emph {et~al.}(2014)\citenamefont
  {O'Shaughnessy}, \citenamefont {Farr}, \citenamefont {Ochsner}, \citenamefont
  {Cho}, \citenamefont {Kim},\ and\ \citenamefont
  {Lee}}]{OShaughnessy:2013zfw}%
  \BibitemOpen
  \bibfield  {author} {\bibinfo {author} {\bibfnamefont {R.}~\bibnamefont
  {O'Shaughnessy}}, \bibinfo {author} {\bibfnamefont {B.}~\bibnamefont {Farr}},
  \bibinfo {author} {\bibfnamefont {E.}~\bibnamefont {Ochsner}}, \bibinfo
  {author} {\bibfnamefont {H.-S.}\ \bibnamefont {Cho}}, \bibinfo {author}
  {\bibfnamefont {C.}~\bibnamefont {Kim}}, \ and\ \bibinfo {author}
  {\bibfnamefont {C.-H.}\ \bibnamefont {Lee}},\ }\href {\doibase
  10.1103/PhysRevD.89.064048} {\bibfield  {journal} {\bibinfo  {journal} {Phys.
  Rev. D}\ }\textbf {\bibinfo {volume} {89}},\ \bibinfo {pages} {064048}
  (\bibinfo {year} {2014})},\ \Eprint {http://arxiv.org/abs/1308.4704}
  {arXiv:1308.4704 [gr-qc]} \BibitemShut {NoStop}%
\bibitem [{\citenamefont {Biscoveanu}\ \emph {et~al.}(2022)\citenamefont
  {Biscoveanu}, \citenamefont {Talbot},\ and\ \citenamefont
  {Vitale}}]{Biscoveanu:2021eht}%
  \BibitemOpen
  \bibfield  {author} {\bibinfo {author} {\bibfnamefont {S.}~\bibnamefont
  {Biscoveanu}}, \bibinfo {author} {\bibfnamefont {C.}~\bibnamefont {Talbot}},
  \ and\ \bibinfo {author} {\bibfnamefont {S.}~\bibnamefont {Vitale}},\ }\href
  {\doibase 10.1093/mnras/stac347} {\bibfield  {journal} {\bibinfo  {journal}
  {Mon. Not. Roy. Astron. Soc.}\ }\textbf {\bibinfo {volume} {511}},\ \bibinfo
  {pages} {4350} (\bibinfo {year} {2022})},\ \Eprint
  {http://arxiv.org/abs/2111.13619} {arXiv:2111.13619 [astro-ph.HE]}
  \BibitemShut {NoStop}%
\bibitem [{\citenamefont {Olsen}\ \emph {et~al.}(2021)\citenamefont {Olsen},
  \citenamefont {Roulet}, \citenamefont {Chia}, \citenamefont {Dai},
  \citenamefont {Venumadhav}, \citenamefont {Zackay},\ and\ \citenamefont
  {Zaldarriaga}}]{Olsen:2021qin}%
  \BibitemOpen
  \bibfield  {author} {\bibinfo {author} {\bibfnamefont {S.}~\bibnamefont
  {Olsen}}, \bibinfo {author} {\bibfnamefont {J.}~\bibnamefont {Roulet}},
  \bibinfo {author} {\bibfnamefont {H.~S.}\ \bibnamefont {Chia}}, \bibinfo
  {author} {\bibfnamefont {L.}~\bibnamefont {Dai}}, \bibinfo {author}
  {\bibfnamefont {T.}~\bibnamefont {Venumadhav}}, \bibinfo {author}
  {\bibfnamefont {B.}~\bibnamefont {Zackay}}, \ and\ \bibinfo {author}
  {\bibfnamefont {M.}~\bibnamefont {Zaldarriaga}},\ }\href {\doibase
  10.1103/PhysRevD.104.083036} {\bibfield  {journal} {\bibinfo  {journal}
  {Phys. Rev. D}\ }\textbf {\bibinfo {volume} {104}},\ \bibinfo {pages}
  {083036} (\bibinfo {year} {2021})},\ \Eprint
  {http://arxiv.org/abs/2106.13821} {arXiv:2106.13821 [astro-ph.HE]}
  \BibitemShut {NoStop}%
\bibitem [{\citenamefont {Chudaykin}\ \emph
  {et~al.}(2021{\natexlab{a}})\citenamefont {Chudaykin}, \citenamefont
  {Dolgikh},\ and\ \citenamefont {Ivanov}}]{Chudaykin:2020ghx}%
  \BibitemOpen
  \bibfield  {author} {\bibinfo {author} {\bibfnamefont {A.}~\bibnamefont
  {Chudaykin}}, \bibinfo {author} {\bibfnamefont {K.}~\bibnamefont {Dolgikh}},
  \ and\ \bibinfo {author} {\bibfnamefont {M.~M.}\ \bibnamefont {Ivanov}},\
  }\href {\doibase 10.1103/PhysRevD.103.023507} {\bibfield  {journal} {\bibinfo
   {journal} {Phys. Rev. D}\ }\textbf {\bibinfo {volume} {103}},\ \bibinfo
  {pages} {023507} (\bibinfo {year} {2021}{\natexlab{a}})},\ \Eprint
  {http://arxiv.org/abs/2009.10106} {arXiv:2009.10106 [astro-ph.CO]}
  \BibitemShut {NoStop}%
\bibitem [{\citenamefont {Chudaykin}\ \emph {et~al.}(2020)\citenamefont
  {Chudaykin}, \citenamefont {Ivanov}, \citenamefont {Philcox},\ and\
  \citenamefont {Simonovi\'c}}]{Chudaykin:2020aoj}%
  \BibitemOpen
  \bibfield  {author} {\bibinfo {author} {\bibfnamefont {A.}~\bibnamefont
  {Chudaykin}}, \bibinfo {author} {\bibfnamefont {M.~M.}\ \bibnamefont
  {Ivanov}}, \bibinfo {author} {\bibfnamefont {O.~H.~E.}\ \bibnamefont
  {Philcox}}, \ and\ \bibinfo {author} {\bibfnamefont {M.}~\bibnamefont
  {Simonovi\'c}},\ }\href {\doibase 10.1103/PhysRevD.102.063533} {\bibfield
  {journal} {\bibinfo  {journal} {Phys. Rev. D}\ }\textbf {\bibinfo {volume}
  {102}},\ \bibinfo {pages} {063533} (\bibinfo {year} {2020})},\ \Eprint
  {http://arxiv.org/abs/2004.10607} {arXiv:2004.10607 [astro-ph.CO]}
  \BibitemShut {NoStop}%
\bibitem [{\citenamefont {Simon}\ \emph {et~al.}(2023)\citenamefont {Simon},
  \citenamefont {Zhang}, \citenamefont {Poulin},\ and\ \citenamefont
  {Smith}}]{Simon:2022lde}%
  \BibitemOpen
  \bibfield  {author} {\bibinfo {author} {\bibfnamefont {T.}~\bibnamefont
  {Simon}}, \bibinfo {author} {\bibfnamefont {P.}~\bibnamefont {Zhang}},
  \bibinfo {author} {\bibfnamefont {V.}~\bibnamefont {Poulin}}, \ and\ \bibinfo
  {author} {\bibfnamefont {T.~L.}\ \bibnamefont {Smith}},\ }\href {\doibase
  10.1103/PhysRevD.107.123530} {\bibfield  {journal} {\bibinfo  {journal}
  {Phys. Rev. D}\ }\textbf {\bibinfo {volume} {107}},\ \bibinfo {pages}
  {123530} (\bibinfo {year} {2023})},\ \Eprint
  {http://arxiv.org/abs/2208.05929} {arXiv:2208.05929 [astro-ph.CO]}
  \BibitemShut {NoStop}%
\bibitem [{\citenamefont {Alam}\ \emph
  {et~al.}(2017{\natexlab{a}})\citenamefont {Alam} \emph
  {et~al.}}]{Alam:2016hwk}%
  \BibitemOpen
  \bibfield  {author} {\bibinfo {author} {\bibfnamefont {S.}~\bibnamefont
  {Alam}} \emph {et~al.} (\bibinfo {collaboration} {BOSS}),\ }\href {\doibase
  10.1093/mnras/stx721} {\bibfield  {journal} {\bibinfo  {journal} {Mon. Not.
  Roy. Astron. Soc.}\ }\textbf {\bibinfo {volume} {470}},\ \bibinfo {pages}
  {2617} (\bibinfo {year} {2017}{\natexlab{a}})},\ \Eprint
  {http://arxiv.org/abs/1607.03155} {arXiv:1607.03155 [astro-ph.CO]}
  \BibitemShut {NoStop}%
%%CITATION = ARXIV:1607.03155;%%
\bibitem [{\citenamefont {Laureijs}\ \emph {et~al.}(2011)\citenamefont
  {Laureijs} \emph {et~al.}}]{Laureijs:2011gra}%
  \BibitemOpen
  \bibfield  {author} {\bibinfo {author} {\bibfnamefont {R.}~\bibnamefont
  {Laureijs}} \emph {et~al.} (\bibinfo {collaboration} {EUCLID}),\ }\href@noop
  {} {\  (\bibinfo {year} {2011})},\ \Eprint {http://arxiv.org/abs/1110.3193}
  {arXiv:1110.3193 [astro-ph.CO]} \BibitemShut {NoStop}%
%%CITATION = ARXIV:1110.3193;%%
\bibitem [{\citenamefont {Aghamousa}\ \emph {et~al.}(2016)\citenamefont
  {Aghamousa} \emph {et~al.}}]{Aghamousa:2016zmz}%
  \BibitemOpen
  \bibfield  {author} {\bibinfo {author} {\bibfnamefont {A.}~\bibnamefont
  {Aghamousa}} \emph {et~al.} (\bibinfo {collaboration} {DESI}),\ }\href@noop
  {} {\  (\bibinfo {year} {2016})},\ \Eprint {http://arxiv.org/abs/1611.00036}
  {arXiv:1611.00036 [astro-ph.IM]} \BibitemShut {NoStop}%
%%CITATION = ARXIV:1611.00036;%%
\bibitem [{\citenamefont {Ivanov}\ \emph {et~al.}(2023)\citenamefont {Ivanov},
  \citenamefont {Philcox}, \citenamefont {Cabass}, \citenamefont {Nishimichi},
  \citenamefont {Simonovi\'c},\ and\ \citenamefont
  {Zaldarriaga}}]{Ivanov:2023qzb}%
  \BibitemOpen
  \bibfield  {author} {\bibinfo {author} {\bibfnamefont {M.~M.}\ \bibnamefont
  {Ivanov}}, \bibinfo {author} {\bibfnamefont {O.~H.~E.}\ \bibnamefont
  {Philcox}}, \bibinfo {author} {\bibfnamefont {G.}~\bibnamefont {Cabass}},
  \bibinfo {author} {\bibfnamefont {T.}~\bibnamefont {Nishimichi}}, \bibinfo
  {author} {\bibfnamefont {M.}~\bibnamefont {Simonovi\'c}}, \ and\ \bibinfo
  {author} {\bibfnamefont {M.}~\bibnamefont {Zaldarriaga}},\ }\href {\doibase
  10.1103/PhysRevD.107.083515} {\bibfield  {journal} {\bibinfo  {journal}
  {Phys. Rev. D}\ }\textbf {\bibinfo {volume} {107}},\ \bibinfo {pages}
  {083515} (\bibinfo {year} {2023})},\ \Eprint
  {http://arxiv.org/abs/2302.04414} {arXiv:2302.04414 [astro-ph.CO]}
  \BibitemShut {NoStop}%
\bibitem [{\citenamefont {Chen}\ \emph {et~al.}(2024)\citenamefont {Chen},
  \citenamefont {Ivanov}, \citenamefont {Philcox},\ and\ \citenamefont
  {Wenzl}}]{Chen:2024vuf}%
  \BibitemOpen
  \bibfield  {author} {\bibinfo {author} {\bibfnamefont {S.-F.}\ \bibnamefont
  {Chen}}, \bibinfo {author} {\bibfnamefont {M.~M.}\ \bibnamefont {Ivanov}},
  \bibinfo {author} {\bibfnamefont {O.~H.~E.}\ \bibnamefont {Philcox}}, \ and\
  \bibinfo {author} {\bibfnamefont {L.}~\bibnamefont {Wenzl}},\ }\href@noop {}
  {\  (\bibinfo {year} {2024})},\ \Eprint {http://arxiv.org/abs/2406.13388}
  {arXiv:2406.13388 [astro-ph.CO]} \BibitemShut {NoStop}%
\bibitem [{\citenamefont {Akitsu}(2024)}]{Akitsu:2024lyt}%
  \BibitemOpen
  \bibfield  {author} {\bibinfo {author} {\bibfnamefont {K.}~\bibnamefont
  {Akitsu}},\ }\href@noop {} {\  (\bibinfo {year} {2024})},\ \Eprint
  {http://arxiv.org/abs/2410.08998} {arXiv:2410.08998 [astro-ph.CO]}
  \BibitemShut {NoStop}%
\bibitem [{\citenamefont {Sullivan}\ \emph {et~al.}(2021)\citenamefont
  {Sullivan}, \citenamefont {Seljak},\ and\ \citenamefont
  {Singh}}]{Sullivan:2021sof}%
  \BibitemOpen
  \bibfield  {author} {\bibinfo {author} {\bibfnamefont {J.~M.}\ \bibnamefont
  {Sullivan}}, \bibinfo {author} {\bibfnamefont {U.}~\bibnamefont {Seljak}}, \
  and\ \bibinfo {author} {\bibfnamefont {S.}~\bibnamefont {Singh}},\ }\href
  {\doibase 10.1088/1475-7516/2021/11/026} {\bibfield  {journal} {\bibinfo
  {journal} {JCAP}\ }\textbf {\bibinfo {volume} {11}},\ \bibinfo {pages} {026}
  (\bibinfo {year} {2021})},\ \Eprint {http://arxiv.org/abs/2104.10676}
  {arXiv:2104.10676 [astro-ph.CO]} \BibitemShut {NoStop}%
\bibitem [{\citenamefont {Baldauf}\ \emph {et~al.}(2016)\citenamefont
  {Baldauf}, \citenamefont {Mirbabayi}, \citenamefont {Simonović},\ and\
  \citenamefont {Zaldarriaga}}]{Baldauf:2016sjb}%
  \BibitemOpen
  \bibfield  {author} {\bibinfo {author} {\bibfnamefont {T.}~\bibnamefont
  {Baldauf}}, \bibinfo {author} {\bibfnamefont {M.}~\bibnamefont {Mirbabayi}},
  \bibinfo {author} {\bibfnamefont {M.}~\bibnamefont {Simonović}}, \ and\
  \bibinfo {author} {\bibfnamefont {M.}~\bibnamefont {Zaldarriaga}},\
  }\href@noop {} {\  (\bibinfo {year} {2016})},\ \Eprint
  {http://arxiv.org/abs/1602.00674} {arXiv:1602.00674 [astro-ph.CO]}
  \BibitemShut {NoStop}%
%%CITATION = ARXIV:1602.00674;%%
\bibitem [{\citenamefont {Chudaykin}\ \emph
  {et~al.}(2021{\natexlab{b}})\citenamefont {Chudaykin}, \citenamefont
  {Ivanov},\ and\ \citenamefont {Simonovi\'c}}]{Chudaykin:2020hbf}%
  \BibitemOpen
  \bibfield  {author} {\bibinfo {author} {\bibfnamefont {A.}~\bibnamefont
  {Chudaykin}}, \bibinfo {author} {\bibfnamefont {M.~M.}\ \bibnamefont
  {Ivanov}}, \ and\ \bibinfo {author} {\bibfnamefont {M.}~\bibnamefont
  {Simonovi\'c}},\ }\href {\doibase 10.1103/PhysRevD.103.043525} {\bibfield
  {journal} {\bibinfo  {journal} {Phys. Rev. D}\ }\textbf {\bibinfo {volume}
  {103}},\ \bibinfo {pages} {043525} (\bibinfo {year} {2021}{\natexlab{b}})},\
  \Eprint {http://arxiv.org/abs/2009.10724} {arXiv:2009.10724 [astro-ph.CO]}
  \BibitemShut {NoStop}%
\bibitem [{\citenamefont {Nishimichi}\ \emph {et~al.}(2020)\citenamefont
  {Nishimichi}, \citenamefont {D'Amico}, \citenamefont {Ivanov}, \citenamefont
  {Senatore}, \citenamefont {Simonovi\'c}, \citenamefont {Takada},
  \citenamefont {Zaldarriaga},\ and\ \citenamefont
  {Zhang}}]{Nishimichi:2020tvu}%
  \BibitemOpen
  \bibfield  {author} {\bibinfo {author} {\bibfnamefont {T.}~\bibnamefont
  {Nishimichi}}, \bibinfo {author} {\bibfnamefont {G.}~\bibnamefont {D'Amico}},
  \bibinfo {author} {\bibfnamefont {M.~M.}\ \bibnamefont {Ivanov}}, \bibinfo
  {author} {\bibfnamefont {L.}~\bibnamefont {Senatore}}, \bibinfo {author}
  {\bibfnamefont {M.}~\bibnamefont {Simonovi\'c}}, \bibinfo {author}
  {\bibfnamefont {M.}~\bibnamefont {Takada}}, \bibinfo {author} {\bibfnamefont
  {M.}~\bibnamefont {Zaldarriaga}}, \ and\ \bibinfo {author} {\bibfnamefont
  {P.}~\bibnamefont {Zhang}},\ }\href {\doibase 10.1103/PhysRevD.102.123541}
  {\bibfield  {journal} {\bibinfo  {journal} {Phys. Rev. D}\ }\textbf {\bibinfo
  {volume} {102}},\ \bibinfo {pages} {123541} (\bibinfo {year} {2020})},\
  \Eprint {http://arxiv.org/abs/2003.08277} {arXiv:2003.08277 [astro-ph.CO]}
  \BibitemShut {NoStop}%
\bibitem [{\citenamefont {Krause}\ \emph {et~al.}(2024)\citenamefont {Krause}
  \emph {et~al.}}]{Beyond-2pt:2024mqz}%
  \BibitemOpen
  \bibfield  {author} {\bibinfo {author} {\bibfnamefont {E.}~\bibnamefont
  {Krause}} \emph {et~al.} (\bibinfo {collaboration} {Beyond-2pt}),\
  }\href@noop {} {\  (\bibinfo {year} {2024})},\ \Eprint
  {http://arxiv.org/abs/2405.02252} {arXiv:2405.02252 [astro-ph.CO]}
  \BibitemShut {NoStop}%
\bibitem [{\citenamefont {D'Amico}\ \emph {et~al.}(2021)\citenamefont
  {D'Amico}, \citenamefont {Senatore},\ and\ \citenamefont
  {Zhang}}]{DAmico:2020kxu}%
  \BibitemOpen
  \bibfield  {author} {\bibinfo {author} {\bibfnamefont {G.}~\bibnamefont
  {D'Amico}}, \bibinfo {author} {\bibfnamefont {L.}~\bibnamefont {Senatore}}, \
  and\ \bibinfo {author} {\bibfnamefont {P.}~\bibnamefont {Zhang}},\ }\href
  {\doibase 10.1088/1475-7516/2021/01/006} {\bibfield  {journal} {\bibinfo
  {journal} {JCAP}\ }\textbf {\bibinfo {volume} {01}},\ \bibinfo {pages} {006}
  (\bibinfo {year} {2021})},\ \Eprint {http://arxiv.org/abs/2003.07956}
  {arXiv:2003.07956 [astro-ph.CO]} \BibitemShut {NoStop}%
\bibitem [{\citenamefont {Hartlap}\ \emph {et~al.}(2007)\citenamefont
  {Hartlap}, \citenamefont {Simon},\ and\ \citenamefont
  {Schneider}}]{Hartlap:2006kj}%
  \BibitemOpen
  \bibfield  {author} {\bibinfo {author} {\bibfnamefont {J.}~\bibnamefont
  {Hartlap}}, \bibinfo {author} {\bibfnamefont {P.}~\bibnamefont {Simon}}, \
  and\ \bibinfo {author} {\bibfnamefont {P.}~\bibnamefont {Schneider}},\ }\href
  {\doibase 10.1051/0004-6361:20066170} {\bibfield  {journal} {\bibinfo
  {journal} {Astron. Astrophys.}\ }\textbf {\bibinfo {volume} {464}},\ \bibinfo
  {pages} {399} (\bibinfo {year} {2007})},\ \Eprint
  {http://arxiv.org/abs/astro-ph/0608064} {arXiv:astro-ph/0608064} \BibitemShut
  {NoStop}%
\bibitem [{\citenamefont {Sellentin}\ and\ \citenamefont
  {Heavens}(2016)}]{Sellentin:2015waz}%
  \BibitemOpen
  \bibfield  {author} {\bibinfo {author} {\bibfnamefont {E.}~\bibnamefont
  {Sellentin}}\ and\ \bibinfo {author} {\bibfnamefont {A.~F.}\ \bibnamefont
  {Heavens}},\ }\href {\doibase 10.1093/mnrasl/slv190} {\bibfield  {journal}
  {\bibinfo  {journal} {Mon. Not. Roy. Astron. Soc.}\ }\textbf {\bibinfo
  {volume} {456}},\ \bibinfo {pages} {L132} (\bibinfo {year} {2016})},\ \Eprint
  {http://arxiv.org/abs/1511.05969} {arXiv:1511.05969 [astro-ph.CO]}
  \BibitemShut {NoStop}%
\bibitem [{\citenamefont {Howlett}\ and\ \citenamefont
  {Percival}(2017)}]{Howlett:2017vwp}%
  \BibitemOpen
  \bibfield  {author} {\bibinfo {author} {\bibfnamefont {C.}~\bibnamefont
  {Howlett}}\ and\ \bibinfo {author} {\bibfnamefont {W.~J.}\ \bibnamefont
  {Percival}},\ }\href {\doibase 10.1093/mnras/stx2342} {\bibfield  {journal}
  {\bibinfo  {journal} {Mon. Not. Roy. Astron. Soc.}\ }\textbf {\bibinfo
  {volume} {472}},\ \bibinfo {pages} {4935} (\bibinfo {year} {2017})},\ \Eprint
  {http://arxiv.org/abs/1709.03057} {arXiv:1709.03057 [astro-ph.CO]}
  \BibitemShut {NoStop}%
%%CITATION = ARXIV:1709.03057;%%
\bibitem [{\citenamefont {Philcox}\ \emph {et~al.}(2021)\citenamefont
  {Philcox}, \citenamefont {Ivanov}, \citenamefont {Zaldarriaga}, \citenamefont
  {Simonovic},\ and\ \citenamefont {Schmittfull}}]{Philcox:2020zyp}%
  \BibitemOpen
  \bibfield  {author} {\bibinfo {author} {\bibfnamefont {O.~H.~E.}\
  \bibnamefont {Philcox}}, \bibinfo {author} {\bibfnamefont {M.~M.}\
  \bibnamefont {Ivanov}}, \bibinfo {author} {\bibfnamefont {M.}~\bibnamefont
  {Zaldarriaga}}, \bibinfo {author} {\bibfnamefont {M.}~\bibnamefont
  {Simonovic}}, \ and\ \bibinfo {author} {\bibfnamefont {M.}~\bibnamefont
  {Schmittfull}},\ }\href {\doibase 10.1103/PhysRevD.103.043508} {\bibfield
  {journal} {\bibinfo  {journal} {Phys. Rev. D}\ }\textbf {\bibinfo {volume}
  {103}},\ \bibinfo {pages} {043508} (\bibinfo {year} {2021})},\ \Eprint
  {http://arxiv.org/abs/2009.03311} {arXiv:2009.03311 [astro-ph.CO]}
  \BibitemShut {NoStop}%
\bibitem [{\citenamefont {Kitaura}\ \emph {et~al.}(2016)\citenamefont {Kitaura}
  \emph {et~al.}}]{Kitaura:2015uqa}%
  \BibitemOpen
  \bibfield  {author} {\bibinfo {author} {\bibfnamefont {F.-S.}\ \bibnamefont
  {Kitaura}} \emph {et~al.},\ }\href {\doibase 10.1093/mnras/stv2826}
  {\bibfield  {journal} {\bibinfo  {journal} {Mon. Not. Roy. Astron. Soc.}\
  }\textbf {\bibinfo {volume} {456}},\ \bibinfo {pages} {4156} (\bibinfo {year}
  {2016})},\ \Eprint {http://arxiv.org/abs/1509.06400} {arXiv:1509.06400
  [astro-ph.CO]} \BibitemShut {NoStop}%
\bibitem [{\citenamefont {Cabass}\ \emph
  {et~al.}(2022{\natexlab{a}})\citenamefont {Cabass}, \citenamefont {Ivanov},
  \citenamefont {Philcox}, \citenamefont {Simonovi\'c},\ and\ \citenamefont
  {Zaldarriaga}}]{Cabass:2022wjy}%
  \BibitemOpen
  \bibfield  {author} {\bibinfo {author} {\bibfnamefont {G.}~\bibnamefont
  {Cabass}}, \bibinfo {author} {\bibfnamefont {M.~M.}\ \bibnamefont {Ivanov}},
  \bibinfo {author} {\bibfnamefont {O.~H.~E.}\ \bibnamefont {Philcox}},
  \bibinfo {author} {\bibfnamefont {M.}~\bibnamefont {Simonovi\'c}}, \ and\
  \bibinfo {author} {\bibfnamefont {M.}~\bibnamefont {Zaldarriaga}},\ }\href
  {\doibase 10.1103/PhysRevLett.129.021301} {\bibfield  {journal} {\bibinfo
  {journal} {Phys. Rev. Lett.}\ }\textbf {\bibinfo {volume} {129}},\ \bibinfo
  {pages} {021301} (\bibinfo {year} {2022}{\natexlab{a}})},\ \Eprint
  {http://arxiv.org/abs/2201.07238} {arXiv:2201.07238 [astro-ph.CO]}
  \BibitemShut {NoStop}%
\bibitem [{\citenamefont {Cabass}\ \emph
  {et~al.}(2022{\natexlab{b}})\citenamefont {Cabass}, \citenamefont {Ivanov},
  \citenamefont {Philcox}, \citenamefont {Simonovi\'c},\ and\ \citenamefont
  {Zaldarriaga}}]{Cabass:2022ymb}%
  \BibitemOpen
  \bibfield  {author} {\bibinfo {author} {\bibfnamefont {G.}~\bibnamefont
  {Cabass}}, \bibinfo {author} {\bibfnamefont {M.~M.}\ \bibnamefont {Ivanov}},
  \bibinfo {author} {\bibfnamefont {O.~H.~E.}\ \bibnamefont {Philcox}},
  \bibinfo {author} {\bibfnamefont {M.}~\bibnamefont {Simonovi\'c}}, \ and\
  \bibinfo {author} {\bibfnamefont {M.}~\bibnamefont {Zaldarriaga}},\ }\href
  {\doibase 10.1103/PhysRevD.106.043506} {\bibfield  {journal} {\bibinfo
  {journal} {Phys. Rev. D}\ }\textbf {\bibinfo {volume} {106}},\ \bibinfo
  {pages} {043506} (\bibinfo {year} {2022}{\natexlab{b}})},\ \Eprint
  {http://arxiv.org/abs/2204.01781} {arXiv:2204.01781 [astro-ph.CO]}
  \BibitemShut {NoStop}%
\bibitem [{\citenamefont {D'Amico}\ \emph {et~al.}(2022)\citenamefont
  {D'Amico}, \citenamefont {Lewandowski}, \citenamefont {Senatore},\ and\
  \citenamefont {Zhang}}]{DAmico:2022gki}%
  \BibitemOpen
  \bibfield  {author} {\bibinfo {author} {\bibfnamefont {G.}~\bibnamefont
  {D'Amico}}, \bibinfo {author} {\bibfnamefont {M.}~\bibnamefont
  {Lewandowski}}, \bibinfo {author} {\bibfnamefont {L.}~\bibnamefont
  {Senatore}}, \ and\ \bibinfo {author} {\bibfnamefont {P.}~\bibnamefont
  {Zhang}},\ }\href@noop {} {\  (\bibinfo {year} {2022})},\ \Eprint
  {http://arxiv.org/abs/2201.11518} {arXiv:2201.11518 [astro-ph.CO]}
  \BibitemShut {NoStop}%
\bibitem [{\citenamefont {Ivanov}\ \emph
  {et~al.}(2020{\natexlab{b}})\citenamefont {Ivanov}, \citenamefont
  {Simonovi\'c},\ and\ \citenamefont {Zaldarriaga}}]{Ivanov:2019hqk}%
  \BibitemOpen
  \bibfield  {author} {\bibinfo {author} {\bibfnamefont {M.~M.}\ \bibnamefont
  {Ivanov}}, \bibinfo {author} {\bibfnamefont {M.}~\bibnamefont {Simonovi\'c}},
  \ and\ \bibinfo {author} {\bibfnamefont {M.}~\bibnamefont {Zaldarriaga}},\
  }\href {\doibase 10.1103/PhysRevD.101.083504} {\bibfield  {journal} {\bibinfo
   {journal} {Phys. Rev. D}\ }\textbf {\bibinfo {volume} {101}},\ \bibinfo
  {pages} {083504} (\bibinfo {year} {2020}{\natexlab{b}})},\ \Eprint
  {http://arxiv.org/abs/1912.08208} {arXiv:1912.08208 [astro-ph.CO]}
  \BibitemShut {NoStop}%
\bibitem [{\citenamefont {Ivanov}\ \emph
  {et~al.}(2020{\natexlab{c}})\citenamefont {Ivanov}, \citenamefont
  {McDonough}, \citenamefont {Hill}, \citenamefont {Simonovi\'c}, \citenamefont
  {Toomey}, \citenamefont {Alexander},\ and\ \citenamefont
  {Zaldarriaga}}]{Ivanov:2020ril}%
  \BibitemOpen
  \bibfield  {author} {\bibinfo {author} {\bibfnamefont {M.~M.}\ \bibnamefont
  {Ivanov}}, \bibinfo {author} {\bibfnamefont {E.}~\bibnamefont {McDonough}},
  \bibinfo {author} {\bibfnamefont {J.~C.}\ \bibnamefont {Hill}}, \bibinfo
  {author} {\bibfnamefont {M.}~\bibnamefont {Simonovi\'c}}, \bibinfo {author}
  {\bibfnamefont {M.~W.}\ \bibnamefont {Toomey}}, \bibinfo {author}
  {\bibfnamefont {S.}~\bibnamefont {Alexander}}, \ and\ \bibinfo {author}
  {\bibfnamefont {M.}~\bibnamefont {Zaldarriaga}},\ }\href {\doibase
  10.1103/PhysRevD.102.103502} {\bibfield  {journal} {\bibinfo  {journal}
  {Phys. Rev. D}\ }\textbf {\bibinfo {volume} {102}},\ \bibinfo {pages}
  {103502} (\bibinfo {year} {2020}{\natexlab{c}})},\ \Eprint
  {http://arxiv.org/abs/2006.11235} {arXiv:2006.11235 [astro-ph.CO]}
  \BibitemShut {NoStop}%
\bibitem [{\citenamefont {Xu}\ \emph {et~al.}(2022)\citenamefont {Xu},
  \citenamefont {Mu\~noz},\ and\ \citenamefont {Dvorkin}}]{Xu:2021rwg}%
  \BibitemOpen
  \bibfield  {author} {\bibinfo {author} {\bibfnamefont {W.~L.}\ \bibnamefont
  {Xu}}, \bibinfo {author} {\bibfnamefont {J.~B.}\ \bibnamefont {Mu\~noz}}, \
  and\ \bibinfo {author} {\bibfnamefont {C.}~\bibnamefont {Dvorkin}},\ }\href
  {\doibase 10.1103/PhysRevD.105.095029} {\bibfield  {journal} {\bibinfo
  {journal} {Phys. Rev. D}\ }\textbf {\bibinfo {volume} {105}},\ \bibinfo
  {pages} {095029} (\bibinfo {year} {2022})},\ \Eprint
  {http://arxiv.org/abs/2107.09664} {arXiv:2107.09664 [astro-ph.CO]}
  \BibitemShut {NoStop}%
\bibitem [{\citenamefont {Nunes}\ \emph {et~al.}(2022)\citenamefont {Nunes},
  \citenamefont {Vagnozzi}, \citenamefont {Kumar}, \citenamefont
  {Di~Valentino},\ and\ \citenamefont {Mena}}]{Nunes:2022bhn}%
  \BibitemOpen
  \bibfield  {author} {\bibinfo {author} {\bibfnamefont {R.~C.}\ \bibnamefont
  {Nunes}}, \bibinfo {author} {\bibfnamefont {S.}~\bibnamefont {Vagnozzi}},
  \bibinfo {author} {\bibfnamefont {S.}~\bibnamefont {Kumar}}, \bibinfo
  {author} {\bibfnamefont {E.}~\bibnamefont {Di~Valentino}}, \ and\ \bibinfo
  {author} {\bibfnamefont {O.}~\bibnamefont {Mena}},\ }\href {\doibase
  10.1103/PhysRevD.105.123506} {\bibfield  {journal} {\bibinfo  {journal}
  {Phys. Rev. D}\ }\textbf {\bibinfo {volume} {105}},\ \bibinfo {pages}
  {123506} (\bibinfo {year} {2022})},\ \Eprint
  {http://arxiv.org/abs/2203.08093} {arXiv:2203.08093 [astro-ph.CO]}
  \BibitemShut {NoStop}%
\bibitem [{\citenamefont {Kumar}\ \emph {et~al.}(2022)\citenamefont {Kumar},
  \citenamefont {Nunes},\ and\ \citenamefont {Yadav}}]{Kumar:2022vee}%
  \BibitemOpen
  \bibfield  {author} {\bibinfo {author} {\bibfnamefont {S.}~\bibnamefont
  {Kumar}}, \bibinfo {author} {\bibfnamefont {R.~C.}\ \bibnamefont {Nunes}}, \
  and\ \bibinfo {author} {\bibfnamefont {P.}~\bibnamefont {Yadav}},\ }\href
  {\doibase 10.1088/1475-7516/2022/09/060} {\bibfield  {journal} {\bibinfo
  {journal} {JCAP}\ }\textbf {\bibinfo {volume} {09}},\ \bibinfo {pages} {060}
  (\bibinfo {year} {2022})},\ \Eprint {http://arxiv.org/abs/2205.04292}
  {arXiv:2205.04292 [astro-ph.CO]} \BibitemShut {NoStop}%
\bibitem [{\citenamefont {Rubira}\ \emph {et~al.}(2023)\citenamefont {Rubira},
  \citenamefont {Mazoun},\ and\ \citenamefont {Garny}}]{Rubira:2022xhb}%
  \BibitemOpen
  \bibfield  {author} {\bibinfo {author} {\bibfnamefont {H.}~\bibnamefont
  {Rubira}}, \bibinfo {author} {\bibfnamefont {A.}~\bibnamefont {Mazoun}}, \
  and\ \bibinfo {author} {\bibfnamefont {M.}~\bibnamefont {Garny}},\ }\href
  {\doibase 10.1088/1475-7516/2023/01/034} {\bibfield  {journal} {\bibinfo
  {journal} {JCAP}\ }\textbf {\bibinfo {volume} {01}},\ \bibinfo {pages} {034}
  (\bibinfo {year} {2023})},\ \Eprint {http://arxiv.org/abs/2209.03974}
  {arXiv:2209.03974 [astro-ph.CO]} \BibitemShut {NoStop}%
\bibitem [{\citenamefont {Rogers}\ \emph {et~al.}(2023)\citenamefont {Rogers},
  \citenamefont {Hlo\v{z}ek}, \citenamefont {Lagu\"e}, \citenamefont {Ivanov},
  \citenamefont {Philcox}, \citenamefont {Cabass}, \citenamefont {Akitsu},\
  and\ \citenamefont {Marsh}}]{Rogers:2023ezo}%
  \BibitemOpen
  \bibfield  {author} {\bibinfo {author} {\bibfnamefont {K.~K.}\ \bibnamefont
  {Rogers}}, \bibinfo {author} {\bibfnamefont {R.}~\bibnamefont {Hlo\v{z}ek}},
  \bibinfo {author} {\bibfnamefont {A.}~\bibnamefont {Lagu\"e}}, \bibinfo
  {author} {\bibfnamefont {M.~M.}\ \bibnamefont {Ivanov}}, \bibinfo {author}
  {\bibfnamefont {O.~H.~E.}\ \bibnamefont {Philcox}}, \bibinfo {author}
  {\bibfnamefont {G.}~\bibnamefont {Cabass}}, \bibinfo {author} {\bibfnamefont
  {K.}~\bibnamefont {Akitsu}}, \ and\ \bibinfo {author} {\bibfnamefont
  {D.~J.~E.}\ \bibnamefont {Marsh}},\ }\href {\doibase
  10.1088/1475-7516/2023/06/023} {\bibfield  {journal} {\bibinfo  {journal}
  {JCAP}\ }\textbf {\bibinfo {volume} {06}},\ \bibinfo {pages} {023} (\bibinfo
  {year} {2023})},\ \Eprint {http://arxiv.org/abs/2301.08361} {arXiv:2301.08361
  [astro-ph.CO]} \BibitemShut {NoStop}%
\bibitem [{\citenamefont {He}\ \emph {et~al.}(2023)\citenamefont {He},
  \citenamefont {Ivanov}, \citenamefont {An},\ and\ \citenamefont
  {Gluscevic}}]{He:2023dbn}%
  \BibitemOpen
  \bibfield  {author} {\bibinfo {author} {\bibfnamefont {A.}~\bibnamefont
  {He}}, \bibinfo {author} {\bibfnamefont {M.~M.}\ \bibnamefont {Ivanov}},
  \bibinfo {author} {\bibfnamefont {R.}~\bibnamefont {An}}, \ and\ \bibinfo
  {author} {\bibfnamefont {V.}~\bibnamefont {Gluscevic}},\ }\href {\doibase
  10.3847/2041-8213/acdb63} {\bibfield  {journal} {\bibinfo  {journal}
  {Astrophys. J. Lett.}\ }\textbf {\bibinfo {volume} {954}},\ \bibinfo {pages}
  {L8} (\bibinfo {year} {2023})},\ \Eprint {http://arxiv.org/abs/2301.08260}
  {arXiv:2301.08260 [astro-ph.CO]} \BibitemShut {NoStop}%
\bibitem [{\citenamefont {He}\ \emph {et~al.}(2024)\citenamefont {He},
  \citenamefont {An}, \citenamefont {Ivanov},\ and\ \citenamefont
  {Gluscevic}}]{He:2023oke}%
  \BibitemOpen
  \bibfield  {author} {\bibinfo {author} {\bibfnamefont {A.}~\bibnamefont
  {He}}, \bibinfo {author} {\bibfnamefont {R.}~\bibnamefont {An}}, \bibinfo
  {author} {\bibfnamefont {M.~M.}\ \bibnamefont {Ivanov}}, \ and\ \bibinfo
  {author} {\bibfnamefont {V.}~\bibnamefont {Gluscevic}},\ }\href {\doibase
  10.1103/PhysRevD.109.103527} {\bibfield  {journal} {\bibinfo  {journal}
  {Phys. Rev. D}\ }\textbf {\bibinfo {volume} {109}},\ \bibinfo {pages}
  {103527} (\bibinfo {year} {2024})},\ \Eprint
  {http://arxiv.org/abs/2309.03956} {arXiv:2309.03956 [astro-ph.CO]}
  \BibitemShut {NoStop}%
\bibitem [{\citenamefont {Camarena}\ \emph {et~al.}(2023)\citenamefont
  {Camarena}, \citenamefont {Cyr-Racine},\ and\ \citenamefont
  {Houghteling}}]{Camarena:2023cku}%
  \BibitemOpen
  \bibfield  {author} {\bibinfo {author} {\bibfnamefont {D.}~\bibnamefont
  {Camarena}}, \bibinfo {author} {\bibfnamefont {F.-Y.}\ \bibnamefont
  {Cyr-Racine}}, \ and\ \bibinfo {author} {\bibfnamefont {J.}~\bibnamefont
  {Houghteling}},\ }\href {\doibase 10.1103/PhysRevD.108.103535} {\bibfield
  {journal} {\bibinfo  {journal} {Phys. Rev. D}\ }\textbf {\bibinfo {volume}
  {108}},\ \bibinfo {pages} {103535} (\bibinfo {year} {2023})},\ \Eprint
  {http://arxiv.org/abs/2309.03941} {arXiv:2309.03941 [astro-ph.CO]}
  \BibitemShut {NoStop}%
\bibitem [{\citenamefont {McDonough}\ \emph {et~al.}(2024)\citenamefont
  {McDonough}, \citenamefont {Hill}, \citenamefont {Ivanov}, \citenamefont
  {La~Posta},\ and\ \citenamefont {Toomey}}]{McDonough:2023qcu}%
  \BibitemOpen
  \bibfield  {author} {\bibinfo {author} {\bibfnamefont {E.}~\bibnamefont
  {McDonough}}, \bibinfo {author} {\bibfnamefont {J.~C.}\ \bibnamefont {Hill}},
  \bibinfo {author} {\bibfnamefont {M.~M.}\ \bibnamefont {Ivanov}}, \bibinfo
  {author} {\bibfnamefont {A.}~\bibnamefont {La~Posta}}, \ and\ \bibinfo
  {author} {\bibfnamefont {M.~W.}\ \bibnamefont {Toomey}},\ }\href {\doibase
  10.1142/S0218271824300039} {\bibfield  {journal} {\bibinfo  {journal} {Int.
  J. Mod. Phys. D}\ }\textbf {\bibinfo {volume} {33}},\ \bibinfo {pages}
  {2430003} (\bibinfo {year} {2024})},\ \Eprint
  {http://arxiv.org/abs/2310.19899} {arXiv:2310.19899 [astro-ph.CO]}
  \BibitemShut {NoStop}%
\bibitem [{\citenamefont {Chudaykin}\ and\ \citenamefont
  {Ivanov}(2019)}]{Chudaykin:2019ock}%
  \BibitemOpen
  \bibfield  {author} {\bibinfo {author} {\bibfnamefont {A.}~\bibnamefont
  {Chudaykin}}\ and\ \bibinfo {author} {\bibfnamefont {M.~M.}\ \bibnamefont
  {Ivanov}},\ }\href {\doibase 10.1088/1475-7516/2019/11/034} {\bibfield
  {journal} {\bibinfo  {journal} {JCAP}\ }\textbf {\bibinfo {volume} {11}},\
  \bibinfo {pages} {034} (\bibinfo {year} {2019})},\ \Eprint
  {http://arxiv.org/abs/1907.06666} {arXiv:1907.06666 [astro-ph.CO]}
  \BibitemShut {NoStop}%
\bibitem [{\citenamefont {Senatore}(2015)}]{Senatore:2014eva}%
  \BibitemOpen
  \bibfield  {author} {\bibinfo {author} {\bibfnamefont {L.}~\bibnamefont
  {Senatore}},\ }\href {\doibase 10.1088/1475-7516/2015/11/007} {\bibfield
  {journal} {\bibinfo  {journal} {JCAP}\ }\textbf {\bibinfo {volume} {1511}},\
  \bibinfo {pages} {007} (\bibinfo {year} {2015})},\ \Eprint
  {http://arxiv.org/abs/1406.7843} {arXiv:1406.7843 [astro-ph.CO]} \BibitemShut
  {NoStop}%
%%CITATION = ARXIV:1406.7843;%%
\bibitem [{\citenamefont {Mirbabayi}\ \emph {et~al.}(2015)\citenamefont
  {Mirbabayi}, \citenamefont {Schmidt},\ and\ \citenamefont
  {Zaldarriaga}}]{Mirbabayi:2014zca}%
  \BibitemOpen
  \bibfield  {author} {\bibinfo {author} {\bibfnamefont {M.}~\bibnamefont
  {Mirbabayi}}, \bibinfo {author} {\bibfnamefont {F.}~\bibnamefont {Schmidt}},
  \ and\ \bibinfo {author} {\bibfnamefont {M.}~\bibnamefont {Zaldarriaga}},\
  }\href {\doibase 10.1088/1475-7516/2015/07/030} {\bibfield  {journal}
  {\bibinfo  {journal} {JCAP}\ }\textbf {\bibinfo {volume} {1507}},\ \bibinfo
  {pages} {030} (\bibinfo {year} {2015})},\ \Eprint
  {http://arxiv.org/abs/1412.5169} {arXiv:1412.5169 [astro-ph.CO]} \BibitemShut
  {NoStop}%
%%CITATION = ARXIV:1412.5169;%%
\bibitem [{\citenamefont {Alam}\ \emph
  {et~al.}(2017{\natexlab{b}})\citenamefont {Alam} \emph
  {et~al.}}]{BOSS:2016wmc}%
  \BibitemOpen
  \bibfield  {author} {\bibinfo {author} {\bibfnamefont {S.}~\bibnamefont
  {Alam}} \emph {et~al.} (\bibinfo {collaboration} {BOSS}),\ }\href {\doibase
  10.1093/mnras/stx721} {\bibfield  {journal} {\bibinfo  {journal} {Mon. Not.
  Roy. Astron. Soc.}\ }\textbf {\bibinfo {volume} {470}},\ \bibinfo {pages}
  {2617} (\bibinfo {year} {2017}{\natexlab{b}})},\ \Eprint
  {http://arxiv.org/abs/1607.03155} {arXiv:1607.03155 [astro-ph.CO]}
  \BibitemShut {NoStop}%
\bibitem [{\citenamefont {Schmittfull}\ \emph {et~al.}(2019)\citenamefont
  {Schmittfull}, \citenamefont {Simonović}, \citenamefont {Assassi},\ and\
  \citenamefont {Zaldarriaga}}]{Schmittfull:2018yuk}%
  \BibitemOpen
  \bibfield  {author} {\bibinfo {author} {\bibfnamefont {M.}~\bibnamefont
  {Schmittfull}}, \bibinfo {author} {\bibfnamefont {M.}~\bibnamefont
  {Simonović}}, \bibinfo {author} {\bibfnamefont {V.}~\bibnamefont {Assassi}},
  \ and\ \bibinfo {author} {\bibfnamefont {M.}~\bibnamefont {Zaldarriaga}},\
  }\href {\doibase 10.1103/PhysRevD.100.043514} {\bibfield  {journal} {\bibinfo
   {journal} {Phys.\ Rev.\ D}\ }\textbf {\bibinfo {volume} {100}},\ \bibinfo
  {pages} {043514} (\bibinfo {year} {2019})},\ \Eprint
  {http://arxiv.org/abs/1811.10640} {arXiv:1811.10640 [astro-ph.CO]}
  \BibitemShut {NoStop}%
\bibitem [{\citenamefont {Schmittfull}\ \emph {et~al.}(2020)\citenamefont
  {Schmittfull}, \citenamefont {Simonovi\'c}, \citenamefont {Ivanov},
  \citenamefont {Philcox},\ and\ \citenamefont
  {Zaldarriaga}}]{Schmittfull:2020trd}%
  \BibitemOpen
  \bibfield  {author} {\bibinfo {author} {\bibfnamefont {M.}~\bibnamefont
  {Schmittfull}}, \bibinfo {author} {\bibfnamefont {M.}~\bibnamefont
  {Simonovi\'c}}, \bibinfo {author} {\bibfnamefont {M.~M.}\ \bibnamefont
  {Ivanov}}, \bibinfo {author} {\bibfnamefont {O.~H.~E.}\ \bibnamefont
  {Philcox}}, \ and\ \bibinfo {author} {\bibfnamefont {M.}~\bibnamefont
  {Zaldarriaga}},\ }\href@noop {} {\  (\bibinfo {year} {2020})},\ \Eprint
  {http://arxiv.org/abs/2012.03334} {arXiv:2012.03334 [astro-ph.CO]}
  \BibitemShut {NoStop}%
\bibitem [{\citenamefont {Taule}\ and\ \citenamefont
  {Garny}(2023)}]{Taule:2023izt}%
  \BibitemOpen
  \bibfield  {author} {\bibinfo {author} {\bibfnamefont {P.}~\bibnamefont
  {Taule}}\ and\ \bibinfo {author} {\bibfnamefont {M.}~\bibnamefont {Garny}},\
  }\href {\doibase 10.1088/1475-7516/2023/11/078} {\bibfield  {journal}
  {\bibinfo  {journal} {JCAP}\ }\textbf {\bibinfo {volume} {11}},\ \bibinfo
  {pages} {078} (\bibinfo {year} {2023})},\ \Eprint
  {http://arxiv.org/abs/2308.07379} {arXiv:2308.07379 [astro-ph.CO]}
  \BibitemShut {NoStop}%
\bibitem [{\citenamefont {Colas}\ \emph {et~al.}(2020)\citenamefont {Colas},
  \citenamefont {D'amico}, \citenamefont {Senatore}, \citenamefont {Zhang},\
  and\ \citenamefont {Beutler}}]{Colas:2019ret}%
  \BibitemOpen
  \bibfield  {author} {\bibinfo {author} {\bibfnamefont {T.}~\bibnamefont
  {Colas}}, \bibinfo {author} {\bibfnamefont {G.}~\bibnamefont {D'amico}},
  \bibinfo {author} {\bibfnamefont {L.}~\bibnamefont {Senatore}}, \bibinfo
  {author} {\bibfnamefont {P.}~\bibnamefont {Zhang}}, \ and\ \bibinfo {author}
  {\bibfnamefont {F.}~\bibnamefont {Beutler}},\ }\href {\doibase
  10.1088/1475-7516/2020/06/001} {\bibfield  {journal} {\bibinfo  {journal}
  {JCAP}\ }\textbf {\bibinfo {volume} {06}},\ \bibinfo {pages} {001} (\bibinfo
  {year} {2020})},\ \Eprint {http://arxiv.org/abs/1909.07951} {arXiv:1909.07951
  [astro-ph.CO]} \BibitemShut {NoStop}%
\bibitem [{\citenamefont {D'Amico}\ \emph {et~al.}(2024)\citenamefont
  {D'Amico}, \citenamefont {Donath}, \citenamefont {Lewandowski}, \citenamefont
  {Senatore},\ and\ \citenamefont {Zhang}}]{DAmico:2022osl}%
  \BibitemOpen
  \bibfield  {author} {\bibinfo {author} {\bibfnamefont {G.}~\bibnamefont
  {D'Amico}}, \bibinfo {author} {\bibfnamefont {Y.}~\bibnamefont {Donath}},
  \bibinfo {author} {\bibfnamefont {M.}~\bibnamefont {Lewandowski}}, \bibinfo
  {author} {\bibfnamefont {L.}~\bibnamefont {Senatore}}, \ and\ \bibinfo
  {author} {\bibfnamefont {P.}~\bibnamefont {Zhang}},\ }\href {\doibase
  10.1088/1475-7516/2024/05/059} {\bibfield  {journal} {\bibinfo  {journal}
  {JCAP}\ }\textbf {\bibinfo {volume} {05}},\ \bibinfo {pages} {059} (\bibinfo
  {year} {2024})},\ \Eprint {http://arxiv.org/abs/2206.08327} {arXiv:2206.08327
  [astro-ph.CO]} \BibitemShut {NoStop}%
\bibitem [{\citenamefont {Ivanov}\ \emph {et~al.}(2021)\citenamefont {Ivanov},
  \citenamefont {Philcox}, \citenamefont {Simonovi\'c}, \citenamefont
  {Zaldarriaga}, \citenamefont {Nishimichi},\ and\ \citenamefont
  {Takada}}]{Ivanov:2021haa}%
  \BibitemOpen
  \bibfield  {author} {\bibinfo {author} {\bibfnamefont {M.~M.}\ \bibnamefont
  {Ivanov}}, \bibinfo {author} {\bibfnamefont {O.~H.~E.}\ \bibnamefont
  {Philcox}}, \bibinfo {author} {\bibfnamefont {M.}~\bibnamefont
  {Simonovi\'c}}, \bibinfo {author} {\bibfnamefont {M.}~\bibnamefont
  {Zaldarriaga}}, \bibinfo {author} {\bibfnamefont {T.}~\bibnamefont
  {Nishimichi}}, \ and\ \bibinfo {author} {\bibfnamefont {M.}~\bibnamefont
  {Takada}},\ }\href@noop {} {\  (\bibinfo {year} {2021})},\ \Eprint
  {http://arxiv.org/abs/2110.00006} {arXiv:2110.00006 [astro-ph.CO]}
  \BibitemShut {NoStop}%
\bibitem [{\citenamefont {Ivanov}\ \emph {et~al.}(2022)\citenamefont {Ivanov},
  \citenamefont {Philcox}, \citenamefont {Nishimichi}, \citenamefont
  {Simonovi\'c}, \citenamefont {Takada},\ and\ \citenamefont
  {Zaldarriaga}}]{Ivanov:2021kcd}%
  \BibitemOpen
  \bibfield  {author} {\bibinfo {author} {\bibfnamefont {M.~M.}\ \bibnamefont
  {Ivanov}}, \bibinfo {author} {\bibfnamefont {O.~H.~E.}\ \bibnamefont
  {Philcox}}, \bibinfo {author} {\bibfnamefont {T.}~\bibnamefont {Nishimichi}},
  \bibinfo {author} {\bibfnamefont {M.}~\bibnamefont {Simonovi\'c}}, \bibinfo
  {author} {\bibfnamefont {M.}~\bibnamefont {Takada}}, \ and\ \bibinfo {author}
  {\bibfnamefont {M.}~\bibnamefont {Zaldarriaga}},\ }\href {\doibase
  10.1103/PhysRevD.105.063512} {\bibfield  {journal} {\bibinfo  {journal}
  {Phys. Rev. D}\ }\textbf {\bibinfo {volume} {105}},\ \bibinfo {pages}
  {063512} (\bibinfo {year} {2022})},\ \Eprint
  {http://arxiv.org/abs/2110.10161} {arXiv:2110.10161 [astro-ph.CO]}
  \BibitemShut {NoStop}%
\bibitem [{\citenamefont {Philcox}\ \emph {et~al.}(2022)\citenamefont
  {Philcox}, \citenamefont {Ivanov}, \citenamefont {Cabass}, \citenamefont
  {Simonovi\'c}, \citenamefont {Zaldarriaga},\ and\ \citenamefont
  {Nishimichi}}]{Philcox:2022frc}%
  \BibitemOpen
  \bibfield  {author} {\bibinfo {author} {\bibfnamefont {O.~H.~E.}\
  \bibnamefont {Philcox}}, \bibinfo {author} {\bibfnamefont {M.~M.}\
  \bibnamefont {Ivanov}}, \bibinfo {author} {\bibfnamefont {G.}~\bibnamefont
  {Cabass}}, \bibinfo {author} {\bibfnamefont {M.}~\bibnamefont {Simonovi\'c}},
  \bibinfo {author} {\bibfnamefont {M.}~\bibnamefont {Zaldarriaga}}, \ and\
  \bibinfo {author} {\bibfnamefont {T.}~\bibnamefont {Nishimichi}},\ }\href
  {\doibase 10.1103/PhysRevD.106.043530} {\bibfield  {journal} {\bibinfo
  {journal} {Phys. Rev. D}\ }\textbf {\bibinfo {volume} {106}},\ \bibinfo
  {pages} {043530} (\bibinfo {year} {2022})},\ \Eprint
  {http://arxiv.org/abs/2206.02800} {arXiv:2206.02800 [astro-ph.CO]}
  \BibitemShut {NoStop}%
\bibitem [{\citenamefont {Herold}\ and\ \citenamefont
  {Ferreira}(2023)}]{Herold:2022iib}%
  \BibitemOpen
  \bibfield  {author} {\bibinfo {author} {\bibfnamefont {L.}~\bibnamefont
  {Herold}}\ and\ \bibinfo {author} {\bibfnamefont {E.~G.~M.}\ \bibnamefont
  {Ferreira}},\ }\href {\doibase 10.1103/PhysRevD.108.043513} {\bibfield
  {journal} {\bibinfo  {journal} {Phys. Rev. D}\ }\textbf {\bibinfo {volume}
  {108}},\ \bibinfo {pages} {043513} (\bibinfo {year} {2023})},\ \Eprint
  {http://arxiv.org/abs/2210.16296} {arXiv:2210.16296 [astro-ph.CO]}
  \BibitemShut {NoStop}%
\bibitem [{\citenamefont {Bakx}\ \emph {et~al.}(2025)\citenamefont {Bakx},
  \citenamefont {Ivanov}, \citenamefont {Philcox},\ and\ \citenamefont
  {Vlah}}]{Bakx:2025pop}%
  \BibitemOpen
  \bibfield  {author} {\bibinfo {author} {\bibfnamefont {T.}~\bibnamefont
  {Bakx}}, \bibinfo {author} {\bibfnamefont {M.~M.}\ \bibnamefont {Ivanov}},
  \bibinfo {author} {\bibfnamefont {O.~H.~E.}\ \bibnamefont {Philcox}}, \ and\
  \bibinfo {author} {\bibfnamefont {Z.}~\bibnamefont {Vlah}},\ }\href@noop {}
  {\  (\bibinfo {year} {2025})},\ \Eprint {http://arxiv.org/abs/2507.22110}
  {arXiv:2507.22110 [astro-ph.CO]} \BibitemShut {NoStop}%
\bibitem [{\citenamefont {Chudaykin}\ \emph {et~al.}(2025)\citenamefont
  {Chudaykin}, \citenamefont {Ivanov},\ and\ \citenamefont
  {Philcox}}]{Chudaykin:2025vdh}%
  \BibitemOpen
  \bibfield  {author} {\bibinfo {author} {\bibfnamefont {A.}~\bibnamefont
  {Chudaykin}}, \bibinfo {author} {\bibfnamefont {M.~M.}\ \bibnamefont
  {Ivanov}}, \ and\ \bibinfo {author} {\bibfnamefont {O.~H.~E.}\ \bibnamefont
  {Philcox}},\ }\href@noop {} {\  (\bibinfo {year} {2025})},\ \Eprint
  {http://arxiv.org/abs/2512.04266} {arXiv:2512.04266 [astro-ph.CO]}
  \BibitemShut {NoStop}%
\end{thebibliography}%

\end{document}